\documentclass[twocolumn]{aastex63}
\usepackage{color}
\usepackage{comment}
\usepackage{inputenc}
\usepackage{import}
\usepackage{CMNDSGen}
\usepackage{CMNDSManual}
\usepackage{CMNDSAngFilt}
\usepackage{CMNDSExt}
\usepackage{CMNDSLXDeltaSZ}
\usepackage{CMNDSNULXPerHost}
\usepackage{CMNDSCont}
\usepackage{CMNDSNumbers}
\usepackage{CMNDSSFRMstar}
\usepackage{CMNDSLXSFRMstarZ}
\usepackage{CMNDSNSOmegaAll}

\shorttitle{Redshift Evolution of ULXs}
\shortauthors{Barrows et al.}

\begin{document}

\accepted{for publication in ApJ}

\title{The Redshift Evolution of Ultraluminous X-Ray Sources out to \emph{z}\,$\sim$\,0.5: Comparison with X-Ray Binary Populations and Contribution to the Cosmic X-Ray Background}

\author[0000-0002-6212-7328]{R. Scott Barrows}
\affiliation{Department of Astrophysical and Planetary Sciences, University of Colorado Boulder, Boulder, CO 80309, USA}

\author{Julia M. Comerford}
\affiliation{Department of Astrophysical and Planetary Sciences, University of Colorado Boulder, Boulder, CO 80309, USA}

\author[0000-0003-2686-9241]{Daniel Stern}
\affiliation{Jet Propulsion Laboratory, California Institute of Technology, 4800 Oak Grove Drive, Pasadena, CA 91109, USA}

\author[0000-0002-1082-7496]{Marianne Heida}
\affiliation{European Southern Observatory, Karl-Schwarzschild-Str. 2, 85748 Garching b. M{\"u}nchen, Germany}

\correspondingauthor{R. Scott Barrows}
\email{Robert.Barrows@Colorado.edu}

\begin{abstract}
\noindent Ultraluminous X-ray sources (ULXs) are thought to be powerful X-ray binaries (XRBs) and may contribute significantly to the redshift-dependent X-ray emission from star forming galaxies. We have assembled a uniform sample of \ULXSZ~ULXs over the redshift range \z\,$=$\,\ZMin\,$-$\,\ZMax~to constrain their physical nature and their contribution to the Cosmic X-Ray Background (CXB). The sample is constructed by crossmatching galaxies from the \sdsstitle~with the \CSC~and selecting off-nuclear X-ray sources after applying astrometric corrections. The fraction of contaminants is $\sim$\,30\%~and shows no evolution with redshift. The host galaxy star formation rates (\SFRs) are systematically elevated relative to the parent sample when matched in host stellar mass. The specific \SFRs~suggest a slight preference for high-mass XRBs, and the X-ray luminosity scaling relations with host galaxy stellar mass and \SFR~indicate that the highest redshift sources represent relatively luminous XRB populations that dominate their host galaxy X-ray emission. The fraction of galaxies hosting at least one ULX of a given luminosity increases with redshift over the full range of our sample, as expected if ULXs are preferentially found in galaxies with high \SFRs~and low metallicities. At \z\,$\sim$\,0.5, the ULX X-ray flux is consistent with the X-ray emission from star-forming galaxies. Moreover, ULXs may account for up to $\sim$\,40\%~of the integrated flux from XRBs in the normal galaxy population out to \z\,$\sim$\,0.5, suggesting they may contribute significantly to the overall ionizing radiation from galaxies.
\end{abstract}


\section{Introduction}
\label{sec:intro}

Ultraluminous X-ray sources (ULXs) are defined as X-ray sources in off-nuclear regions of galaxies with observed fluxes that (assuming isotropic emission) correspond to luminosities exceeding the theoretical Eddington limit for accretion onto stellar mass compact objects \citep[for a review see][]{Kaaret:2017}. The adopted lower luminosity limits of ULX categorization vary from \LX\,$=$\,$2\times10^{38}$\,\uLum~(Eddington limit for a 1.4\,\Msun~neutron star) to \LX\,$=$\,$3\times10^{39}$\,\uLum~(Eddington limit for a $\sim$\,20\,\Msun~black hole; BH). Super-Eddington accretion onto stellar remnants is considered the most likely explanation for ULXs \citep[e.g.][]{King:2001,Gladstone:2009,Sutton:2013} and has been confirmed in several nearby cases via neutron star pulsed X-ray emission \citep{Bachetti:2014,Furst:2016,Israel:2017,Israel:2017b} and resonance features due to a magnetic field \citep{Brightman:2018}. A significant fraction of other well-studied nearby ULXs observed by the \emph{Nuclear Spectroscopic Telescope Array} show hard X-ray excesses that may also be due to pulsation \citep[e.g.][]{Walton:2013,Bachetti:2013,Walton:2015,Mukherjee:2015,Luangtip:2016,Walton:2018b}, and super-Eddington accretion onto a neutron star has been observed to produce an X-ray luminosity of up to \LX\,$\sim$\,$2\times10^{41}$\,\uLum~\citep{Israel:2017}. On the other hand, more luminous off-nuclear X-ray sources are referred to as hyperluminous X-ray sources (HLXs) and are more likely associated with accretion onto intermediate or super-massive BHs (IMBHs and SMBHs, respectively) with masses of \MBH\,$>$1,000\,\Msun~(e.g. \citealt{King:2005}; for a review see \citealt{Mezcua:2017}).

If ULXs are powered by accretion onto stellar remnants, then they are likely a subset of the X-ray binary (XRB) population in which the accretors are either neutron stars or BHs in a gravitationally-bound system with a donor star that supplies the accreted mass \citep[for a review of Galactic XRBs see][]{Remillard:McClintock:2006}. XRBs with donor star masses of $<$($>$) 10\,\Msun~are referred to as low(high)-mass XRBs (LMXBs and HMXBs, respectively), and the mass is transferred from the donor star via Roche-lobe overflow or from stellar winds. Given that massive stars can provide more material for accretion, the most luminous ULXs are more likely powered by HMXBs with mass-transfer rates that can exceed the Eddington limit \citep[e.g.][]{Pavlovskii:2017}, though geometrical beaming along the line-of-sight can significantly augment the observed luminosities \citep[e.g.][]{Middleton:2017}.

The long-lived donor stars of LMXBs mean that their global emissivity is most strongly correlated with their host galaxy total stellar masses \citep[e.g.][]{Gilfanov:2004,Boroson:2011}. On the other hand, the emissivity of HMXBs is tied to their host galaxy star formation rates (\SFRs) that trace the formation of massive and short-lived donor stars \citep[e.g.][]{Grimm:2003,Ranalli:2003,Hornschemeier:2005,Mineo:2012}. Since the total stellar masses and \SFRs~of galaxies evolve with redshift \citep[e.g.][and references therein]{Madau:2014}, and their metallicities may have a significant impact on the form of this evolution \citep[e.g.][]{Fornasini:2019,Fornasini:2020}, the formation frequency and nature of XRBs will also evolve with redshift. This evolution has been empirically constrained out to high redshifts through X-ray stacking \citep[e.g.][]{Basu-Zych:2013,Lehmer:2016,Aird:2017}, and comparison with stellar population synthesis models \citep{Fragos:2013,Madau:2017} suggests that HMXBs dominate the XRB emissivity at high redshifts, while LMXBs dominate after \z\,$\sim$\,2 due to increasing galaxy metallicities. Moreover, aggregate XRB populations produce most of the X-ray emission from normal (i.e. without active galactic nuclei; AGN) galaxies \citep[e.g.][]{Lehmer:2012} and may contribute up to $\sim$\,20\%~of the Cosmic X-ray Background (CXB) when integrated out to $z=10$ \citep[e.g.][]{Dijkstra:2012}.

However, whether or not ULXs follow a similar evolutionary path is currently not clear. Indeed, ULXs dominate the point source luminosities of normal star-forming galaxies \citep[e.g.][]{Colbert:2004,Fabbiano:2006}, and the CXB may therefore be affected by the redshift evolution of ULXs. This contribution has strong implications for several areas of modern astrophysics, including constraints on the sources of interstellar medium (ISM) heating and feedback \citep[e.g.][]{Pakull:2010,Soria:2014,Lopez:2019}, empirical laboratories for extreme mass accretion rates and/or geometrical beaming, and for the progenitors of gravitational wave sources produced by the coalescence of BHs more massive than known Galactic stellar remnants \citep[e.g.][]{Abbott:2016,Nitz:2020}. 

Furthermore, some ULXs may alternatively be low luminosity AGN in the tidally-stripped cores of galaxies that merged with the host galaxy and are wandering within its gravitational potential \citep[e.g.][]{Farrell:2009,Comerford:2015}. Given that ULX optical counterparts are typically much fainter than expected for the stripped bulges of massive galaxies, in the AGN scenario they likely represent IMBHs from dwarf galaxies \citep[e.g.][]{King:2005,Wolter:2006,Feng:Kaaret:2009,Jonker:2010,Mezcua:2015,Barrows:2019}. The IMBH scenario may also be possible if it formed in-situ within a dense stellar system such as a globular cluster \citep[e.g.][]{Sigurdsson:1993,Miller:2002,Maccarone:2007} or a compact star cluster \citep[e.g.][]{Ebisuzaki:2001}.

While ULXs in the nearby Universe have been studied extensively \citep{Colbert:2002,Swartz:2004,Liu:Bregman:2005,Liu:2005,Liu.J:2011,Swartz:2011,Walton:2011,Gong:2016,Earnshaw:2019,Kovlakas:2020,Inoue:2021,Walton:2022}, their properties past \z\,$\sim$\,0.05 are poorly constrained. \citet{Hornschemeier:2004} originally identified 10 intermediate-redshift ULX candidates\footnote{The term `candidate' is used because a fraction of ULXs are expected to be unrelated background or foreground sources. Spectroscopic redshifts of the accreting sources are the best means of confirming or rejecting association with the host galaxy.} from the \ch~Deep Field$-$North and the \ch~Deep Field$-$South (\z\,$=$\,0.038\,$-$\,0.232, with a median of \z\,$\sim$\,0.11). \citet{Lehmer:2006} identified 15 additional ULX candidates by also incorporating the Extended \ch~Deep Field$-$South (\z\,$=$\,0.038\,$-$\,0.298, with a median of \z\,$\sim$\,0.14). Both studies found tentative evidence that the fraction of galaxies hosting ULXs is larger at intermediate redshifts compared to locally. \citet{Mainieri:2010} subsequently identified 7 new ULX candidates in the Cosmic Evolution Survey \citep[COSMOS;][]{Scoville:2007} field (\z\,$=$\,0.072\,$-$\,0.283, with a median of \z\,$\sim$\,0.13), finding a smaller occupation fraction that suggests a weaker redshift dependence. In this paper we build upon these results by studying the redshift evolution of ULXs using a large and uniformly-constructed sample that ranges from the local Universe out to intermediate redshifts (\z\,$\sim$\,0.002\,$-$\,0.5) for the first time.

This paper is structured as follows: in Section \ref{sec:sample} we describe the steps taken to create the sample, in Section \ref{sec:cont} we estimate the fraction of unknown background or foreground contaminating X-ray sources, in Section \ref{sec:sfr_mstar} we estimate the host galaxy properties, in Section \ref{sec:com_xrb} we compare the ULXs with XRB populations, in Section \ref{sec:frac} we examine the ULX occupation fraction, in Section \ref{sec:cxb} we determine the contribution of ULXs to the CXB, and in Section \ref{sec:conc} we present our conclusions. Throughout we assume a flat cosmology defined by the nine-year Wilkinson Microwave Anisotropy Probe observations \citep{Hinshaw:2013}: \HNaught$\,=\,$\HNaughtValue~\uHNaught~and \OmegaM$\,=\,$\OmegaMValue.

\section{Building the Sample}
\label{sec:sample}

Our procedure for building the sample of ULX candidates is as follows: selection of the initial galaxy sample (Section \ref{sec:galselect}) and the initial X-ray source sample (Section \ref{sec:xrayselect}), spatial cross-match of the galaxies and X-ray sources (Section \ref{sec:initmatch}), selection of spatially offset X-ray sources (Section \ref{sec:regastr}), application of X-ray luminosity thresholds that target ULXs (Section \ref{sec:xrayspec}), and removal of AGN and known contaminants (Section \ref{sec:AGN}). Basic properties of the final sample are summarized in Section \ref{sec:final}, and the effects of source confusion are discussed in Section \ref{sec:confusion}.

\subsection{Galaxy Selection}
\label{sec:galselect}

The initial galaxy sample is derived from the catalog of \sdsstitle~(\sdss) detections in Data Release 16 \citep[DR16;][]{Ahumada:2020} that are classified as \texttt{Galaxy} and with measured photometric redshifts (\photoz). We remove all detections that were flagged as saturated by the SDSS pipeline. If a galaxy is in the \sdss~spectroscopic sample, then we adopt the spectroscopic redshift (\specz) as the final redshift value (\z). Otherwise, we query the \nedname~(\ned) for a spectroscopic redshift using a crossmatch radius of $5''$. If multiple source matches are found, we take the closest match. Furthermore, if a matched source has multiple values of \specz, then the value with the smallest uncertainty is used.

\begin{figure}[t!]
\includegraphics[width=0.475\textwidth]{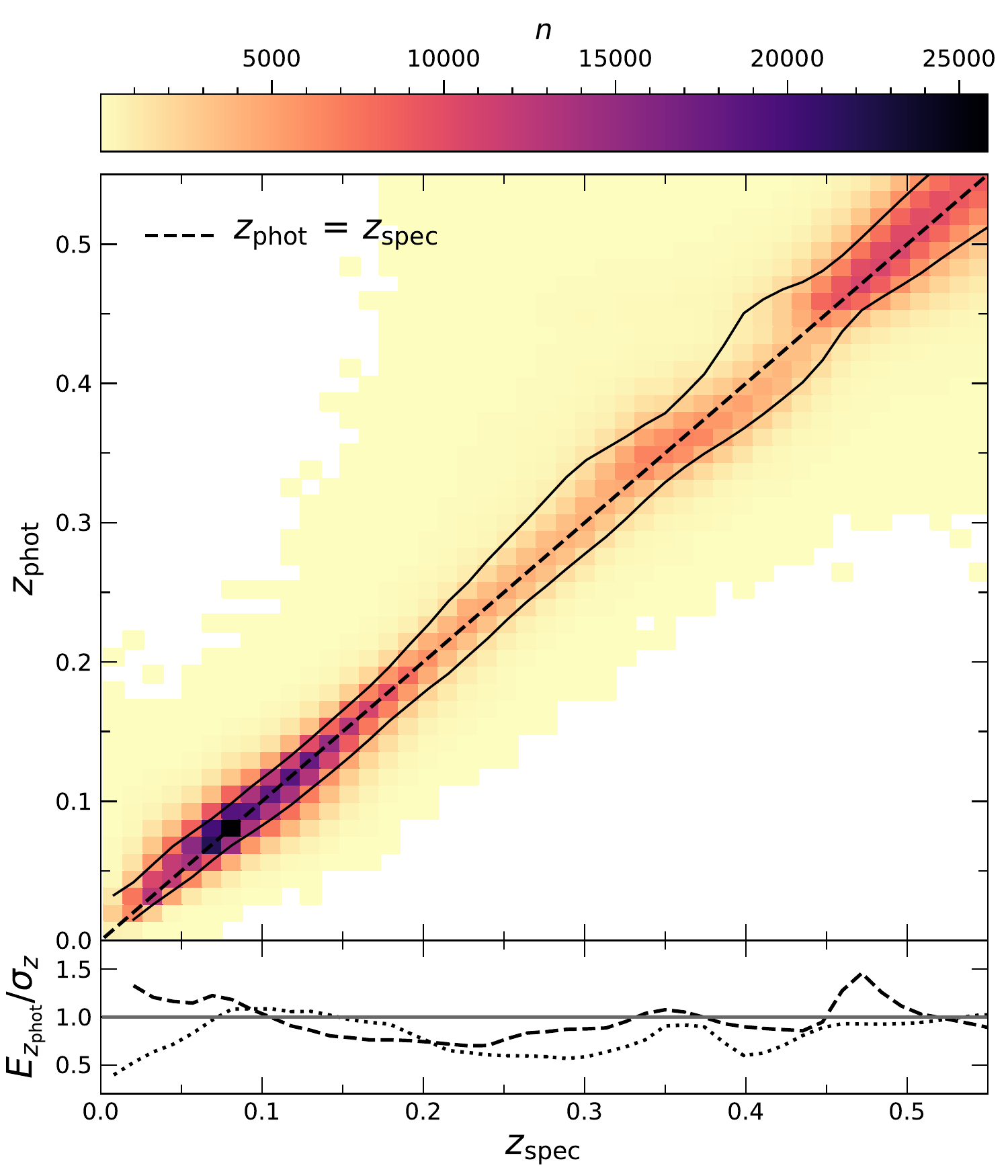}
\vspace{-0.2in}
\caption{\footnotesize{Top: Photometric redshifts (\photoz) against spectroscopic redshifts (\specz) for the subset of the initial galaxy sample (Section \ref{sec:galselect}) with \specz~values. The plotted redshift range encompasses our final sample (Section \ref{sec:final}). Number densities ($n$) are shown for $n$\,$>$\,20. The solid lines bound the 68.3\% confidence lower and upper intervals ($\sigma_{\mathrm{z,lo}}$ and $\sigma_{\mathrm{z,hi}}$, respectively) around the one-to-one relation (dashed). Bottom: ratio of the photometric redshift error (\photozerr) to $\sigma_{\mathrm{z,lo}}$ (dashed) and to $\sigma_{\mathrm{z,hi}}$ (dotted). The horizontal line indicates \photozerr\,$=$\,$\sigma_{\mathrm{z}}$. Both $\sigma_{\mathrm{z,lo}}$ and $\sigma_{\mathrm{z,hi}}$ are overall accurately traced by \photozerr~though they are on average larger. Therefore, we adopt $\sigma_{\mathrm{z,lo}}$ and $\sigma_{\mathrm{z,hi}}$ as our final \photoz~errors.}}
\label{fig:zspec_zphot}
\end{figure}

If no value of \specz~is available, then that of \photoz~is used. Values of \photoz~were derived by the \sdss~pipeline based on a training sample of galaxies with spectroscopic redshifts and with similar colors and $r$-band magnitudes\footnote{\href{https://www.sdss.org/dr12/algorithms/photo-z/}{https://www.sdss.org/dr12/algorithms/photo-z/}}. The \photoz~accuracy of the parent galaxy sample (as quantified by comparison to the \specz~values) is illustrated in the top panel of Figure \ref{fig:zspec_zphot}. The bottom panel of Figure \ref{fig:zspec_zphot} demonstrates how the photometric redshift errors \citep[\photozerr; described in][]{Scranton:2005} are generally reliable tracers of the true accuracy as quantified by the lower and upper 68.3\%~bounds around the one-to-one relation ($\sigma_{\mathrm{z,lo}}$ and $\sigma_{\mathrm{z,hi}}$, respectively). However, on average the \photozerr~values are smaller than both $\sigma_{\mathrm{z,lo}}$ and $\sigma_{\mathrm{z,hi}}$. Therefore, to avoid systematic under-estimates of \photozerr, for the \photoz~errors we use $\sigma_{\mathrm{z,lo}}$ and $\sigma_{\mathrm{z,hi}}$.

\subsection{X-Ray Source Selection}
\label{sec:xrayselect}

The High Resolution Camera (HRC) and the Advanced CCD Imaging Spectrometer (ACIS) on the \ch~\emph{X-Ray Observatory} provide the best spatial resolution of current X-ray telescopes \citep{Weisskopf:2000} and are therefore optimal for identifying off-nuclear X-ray sources out to intermediate redshifts. To obtain the most comprehensive list of robust source detections from \ch, we use the \ch~Source Catalog \citep[][]{Evans:2010} Version 2 (\CSCTwo) Master Sources as our initial sample of X-ray sources.

Since ULXs are defined as point sources, we omit X-ray sources with a 68\%~lower confidence limit on the $1\sigma$ major axis extent that is greater than the point spread function (PSF) $1\sigma$ radius at the source position. Since each Master Source from the \CSCTwo~can be in multiple observations (OBSIDs), we compute the PSF as the mean value obtained from the PSF maps of each OBSID in the `best' Bayesian block of observations\footnote{Each Bayesian block contains observations with a constant photon flux \citep{Scargle:2013}, and the block with the largest combined exposure time is the `best' block: \href{http://cxc.harvard.edu/csc2/data\_products/master/blocks3.html}{http://cxc.harvard.edu/csc2/data\_products/master/blocks3.html}}. We further remove any Master Sources flagged as extended (\texttt{extent\_flag\,=\,TRUE}) by the \CSCTwo~pipeline\footnote{\href{https://cxc.harvard.edu/csc/columns/flags.html}{https://cxc.harvard.edu/csc/columns/flags.html}}.

\begin{figure}[t!]
\includegraphics[width=0.475\textwidth]{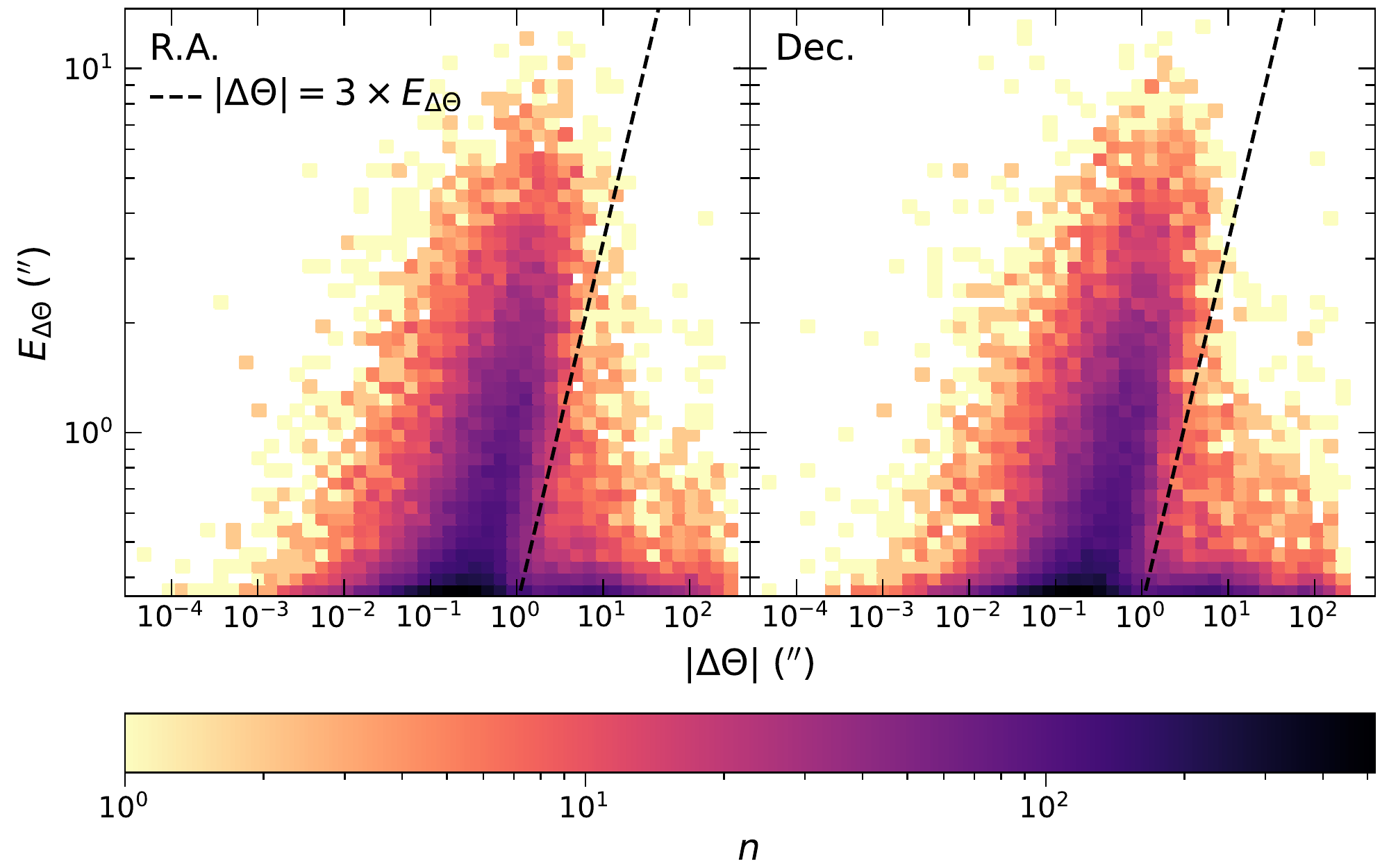}
\vspace{-0.2in}
\caption{\footnotesize{Angular offset uncertainty between the X-ray source and the galaxy centroid ($E_{\Delta \Theta}$) against the angular offset ($\Delta \Theta$) along the right ascension (left) and declination (right) dimensions for the sample of matched galaxy and X-ray source pairs (Section \ref{sec:initmatch}). The dashed line indicates $\Delta \Theta=3\times E_{\Delta \Theta}$ (angular offset criteria used for ULX candidate selection; Section \ref{sec:regastr}).}}
\label{fig:DELTA_S_DELTA_THETA_PLOT}
\end{figure}

\begin{figure}[t!]
\includegraphics[width=0.475\textwidth]{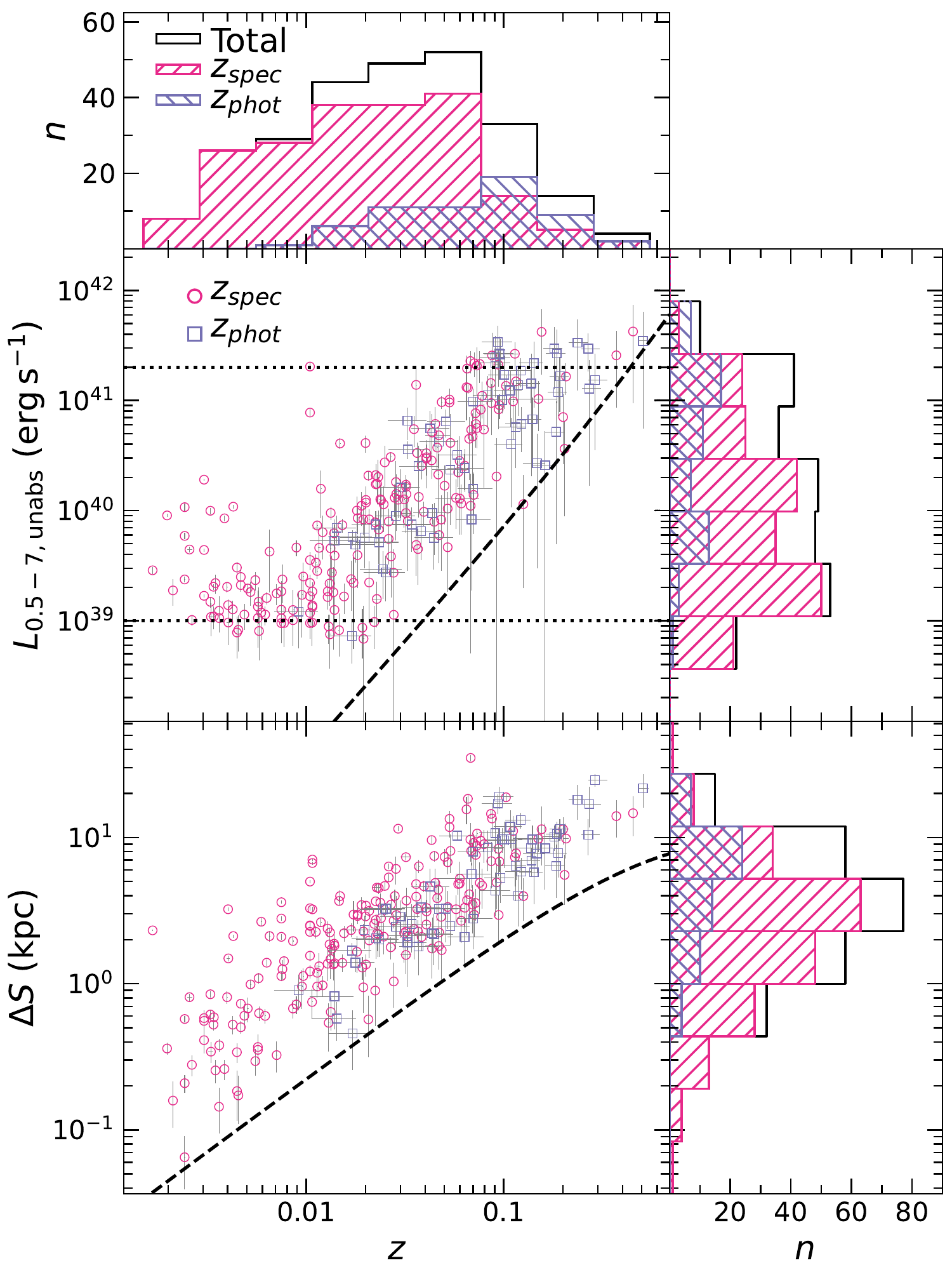}
\vspace{-0.2in}
\caption{\footnotesize{Unabsorbed, rest-frame 0.5$-$7\,keV luminosity (\LXBUnabs; top) and projected physical offset from the host galaxy centroid (\DeltaS; bottom) against redshift (\z) for our final sample of ULX candidates (Section \ref{sec:final}). The samples with and without spectroscopic redshifts are indicated by the magenta circles and purple squares, respectively. The positive correlations of both \LXBUnabs~and \DeltaS~with \z~result in redshift-dependent selection biases toward luminous ULX candidates with large physical offsets at high redshifts. The dashed lines represent the minimum 0.5$-$7\,keV flux sensitivity (top) and the minimum resolvable physical offset (bottom) from the final sample. The dotted lines denote the lower and upper luminosity thresholds for ULX selection (Section \ref{sec:xrayspec}).}}
\label{fig:LX_DELTA_S_Z_PLOT}
\end{figure}

\subsection{Matching X-Ray Sources to Galaxies}
\label{sec:initmatch}

The parent galaxy sample is the subset of the initial galaxy sample (Section \ref{sec:galselect}) that is within the \CSCTwo~footprint. Matches between the parent galaxy sample and the X-ray source sample (Section \ref{sec:xrayselect}) are based on their world coordinates. The \sdss~galaxy positions are defined by the \rband-band photometric centroids, and the coordinates of each X-ray source are based on the \CSCTwo~Maximum Likelihood Estimation\footnote{\href{https://cxc.harvard.edu/csc/columns/positions.html}{https://cxc.harvard.edu/csc/columns/positions.html}}.

We require that each X-ray source centroid is within one Petrosian radius (\rpetro; measured by the \sdss~pipeline) of a galaxy centroid. We remove any galaxies with Petrosian magnitudes fainter than $r$\,$=$\,21 as the Petrosian radii become significantly less accurate above that threshold. Multiple X-ray sources may satisfy this criterion for a single galaxy. Even though radii of 2\,$\times$\,\rpetro~provide the optimal combination of maximizing the integrated galaxy flux while minimizing sky noise \citep[e.g.][]{Graham:2005b}, we observe that X-ray sources with angular offsets $>$\,1\,$\times$\,\rpetro~are significantly more likely to have optical counterparts detected in the \sdss~imaging (see Section \ref{sec:AGN}). Since these optical detections are likely associated with external galaxies (either interacting or background galaxies) and hence do not satisfy the traditional definition of ULXs, we retain the upper angular offset threshold of 1\,$\times$\,\rpetro~to exclude them. This procedure yields \PetroAngFiltSZ~unique matches between X-ray sources and galaxies.

While the Petrosian radii do not account for the apparent elliptical profiles of inclined galaxies, they are more robust than the \sdss~exponential model ellipticities, particularly for the fainter galaxies in the parent galaxy sample. However, to quantify the impact of galaxy inclinations on our selection, we scale the exponential model major and minor radii to yield an elliptical area that equals the circular area defined by the Petrosian radius. The number of ULX candidates selected based on the elliptical galaxy profiles is 97\% of the number selected based on the circular galaxy profiles. Furthermore, out of our sample selected using the Petrosian radius, 95\% would be selected using the corresponding elliptical profiles. We correct our estimates of ULX occupation fraction (Section \ref{sec:frac}) and CXB contribution (Section \ref{sec:cxb}) for the 5\%~that may be outside of the galaxy profile ellipse.

\subsection{Selection of Spatially Offset X-Ray Sources}
\label{sec:regastr} 

To obtain estimates of the relative astrometric accuracy between the galaxy and X-ray source positions we follow the procedure outlined in \citet{Barrows:2016,Barrows:2019}. Here we reiterate the basic steps: we first identify significantly-detected sources ($>$3$\sigma$) in the SDSS \rband-band images using \se~\citep{Bertin:Arnouts:1996} and in the \ch~images using \wvd~(as part of the \ciaotitle~software; \ciao) and a probability threshold of $10^{-8}$. We then filter out unreliable \sdss~detections (sources at frame edges and blended sources) and extended sources from both source lists. The host galaxies and candidate off-nuclear X-ray sources are excluded from the source lists to produce astrometric corrections that are independent of the spatial offsets being tested. Matched pairs of sources between the \se~and \wvd~lists are identified within a $2''$ threshold radius.

Translational corrections along the right ascension and declination are computed as the mean offset between the final matched source lists after iteratively rejecting matched pairs that are outliers by more than $1.5\sigma$. While astrometric corrections that include a term accounting for rotation and scale factors are more general, they require a large number of matched pairs that are distributed evenly throughout the images for accurate solutions. Since this is not possible for the majority of our sample (the median number of matched pairs is three), for uniformity we do not include this extra term when computing corrections. To quantify the impact of this choice on our results, for the subset with four or more matched pairs we compare the translation-only corrections to those that include a rotation and scale factor term (computed using \texttt{wcs\-match} within \ciao). We find that the differences in the derived X-ray source positional corrections are negligible and have no effect on the specific ULX candidates selected.

\begin{figure}[t!]
\includegraphics[width=0.475\textwidth]{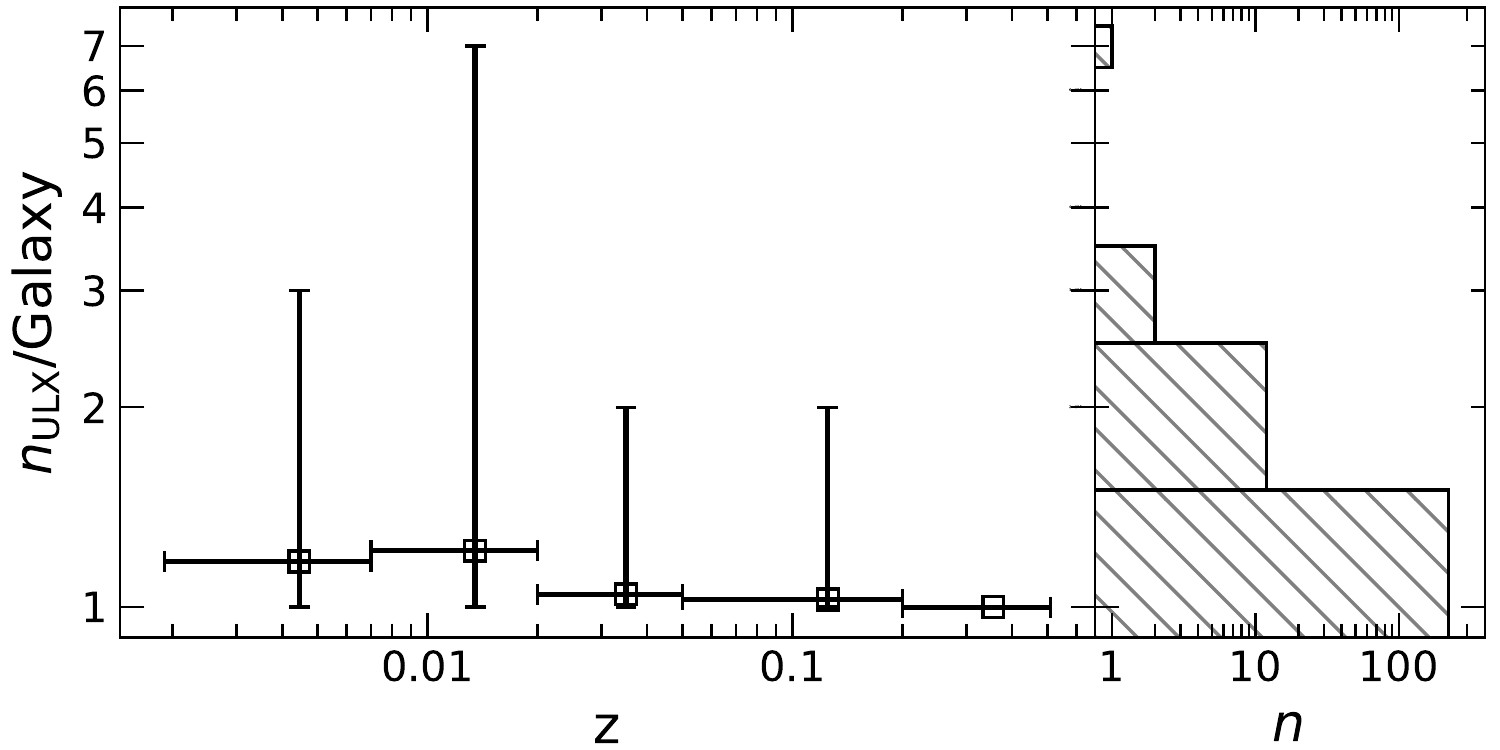}
\vspace{-0.2in}
\caption{\footnotesize{Number of ULX candidates in each host galaxy ($n_{\mathrm{ULX}}/\mathrm{Galaxy}$) against the host galaxy redshift (\z). Data points represent mean values in redshift bins that are approximately even in logarithmic-space and adjusted to have a minimum of fifteen ULX candidates per bin: \z\,$=$\,0.002\,$-$\,0.007, 0.007\,$-$\,0.02, 0.02\,$-$\,0.05, 0.05\,$-$\,0.2, and 0.2\,$-$\,0.51. Vertical errorbars represent the full range of values in each bin, and horizontal errorbars denote the bin width. The distribution of $n_{\mathrm{ULX}}$ is shown on the right. $n_{\mathrm{ULX}}$ decreases with redshift due to the decreasing angular size of the host galaxies and the \CSCTwo~sensitivity limits.}}
\label{fig:NULX_PER_GAL_PLOT}
\end{figure}

The relative astrometric uncertainties are computed from the quadrature sum of the errors on the source centroids in the final matched list. If no matches are found between a pair of images, then the translational corrections are set to zero, and the astrometric uncertainties are set to the quadrature sum of the absolute astrometric errors from the \sdss~($0\farcs035$) and \ch~($0\farcs8$). 

\begin{figure*}[t!]
\includegraphics[width=\textwidth]{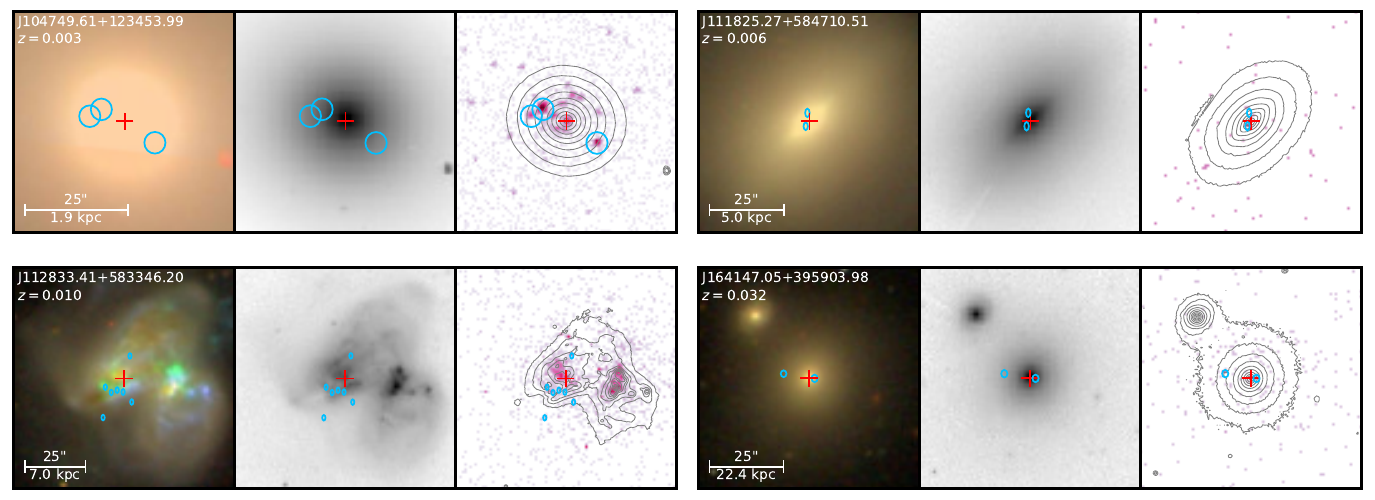}
\vspace{-0.2in}
\caption{\footnotesize{Host galaxies with multiple ULX candidates. Left: \sdss~$g+r+i$ color composite image; middle: \pnstrs~\iband-band image; right: \ch~0.5$-$7\,keV rest-frame image with the \pnstrs~\iband-band image contours overlaid. The galaxy centroid is marked by a red cross and the ULX candidate positions and errors are indicated with blue ellipses. The examples represent host galaxies with the maximum number of ULX candidates in redshift bins of approximately even logarithmic spacing over the interval \z\,$=$\,0.002\,$-$\,0.06 and defined by the following boundaries: \z\,$=$\,[0.002, 0.005, 0.01, 0.03, 0.06].}}
\label{fig:overlays_multiple}
\end{figure*}

The transformations are computed between the \sdss~\rband-band image and each \ch~OBSID in which the X-ray source is detected, and the final corrections between a galaxy and X-ray source position are the error-weighted averages of the astrometric corrections between each of the individual image pairs. These corrections are then applied to each X-ray source to put them in the \sdss~reference frame. The uncertainties on those final transformations are the standard error of the weighted mean. Then we reapply the step requiring corrected X-ray source positions to be within one Petrosian radius of the galaxy centroid (Section \ref{sec:initmatch}). 

The uncertainty of the X-ray source position relative to the galaxy centroid is the quadrature sum of the final relative astrometric uncertainties, the X-ray source centroid uncertainty, and the galaxy centroid uncertainty. These uncertainties correspond to $1\sigma$ confidence intervals and are computed separately for both the right ascension and declination. Spatially offset X-ray sources are selected as being offset from the host galaxy centroid by $\geq$\,3 times the offset uncertainty along either the right ascension or declination (Figure \ref{fig:DELTA_S_DELTA_THETA_PLOT}). 

We also remove galaxies for which offsets can not be reliably measured. These are based on a visual inspection and consist of galaxies for which the \sdss~photometric detection is not located at the galaxy nucleus, galaxies with dust lanes that may affect the detection of the nucleus, or galaxies with photometry that may be contaminated by bright neighboring sources. After this procedure, we are left with \AngFiltSZ~X-ray sources that are spatially offset from their candidate host galaxy centroid.

\begin{figure*}[t!]
\includegraphics[width=\textwidth]{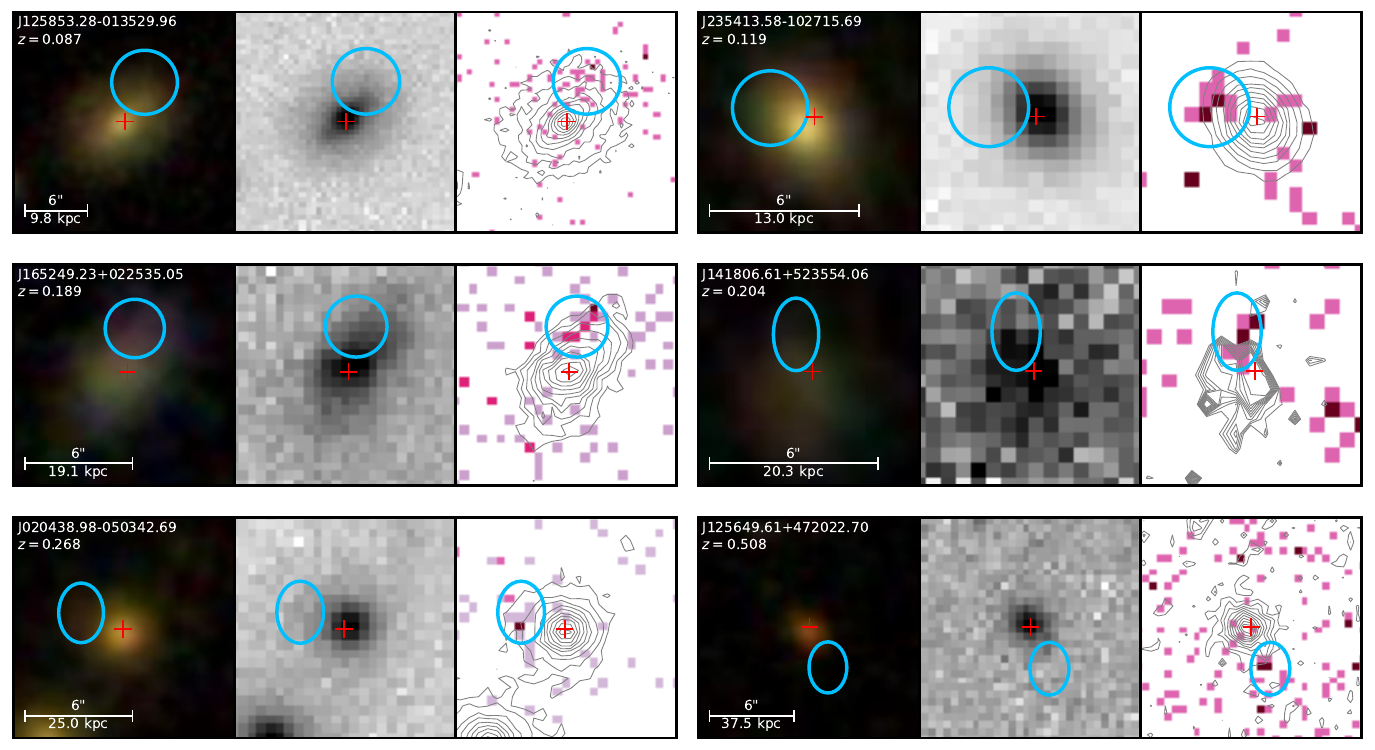}
\vspace{-0.2in}
\caption{\footnotesize{Same as Figure \ref{fig:overlays_multiple} but for host galaxies over the redshift range not sampled by galaxies with more than one ULX. The examples represent ULX candidates with the median \LXBUnabs~value in redshift bins of approximately even logarithmic spacing over the interval \,z\,$=$\,0.06\,$-$\,0.52 and defined by the following boundaries: \z\,$=$\,[0.06, 0.1, 0.15, 0.20, 0.25, 0.35, 0.52].}}
\label{fig:overlays}
\end{figure*}

\subsection{X-Ray Spectral Modeling and Application of Luminosity Thresholds}
\label{sec:xrayspec}

We convert the observed \CSCTwo~0.5$-$7\,keV aperture fluxes to unabsorbed, rest-frame 0.5$-$7\,keV fluxes assuming an intrinsic model for the accreting source. Based on detailed spectral modeling of nearby ULXs \citep[e.g.][]{Swartz:2004,Winter:2006,Gladstone:2009,Sutton:2013,Walton:2018b}, the X-ray emission is often successfully described by a combination of a blackbody component and a powerlaw component. While most \CSCTwo~sources do not have sufficient counts for multi-component spectral models, powerlaw and blackbody models are provided by the \CSCTwo \footnote{\href{https://cxc.cfa.harvard.edu/csc/columns/spectral\_properties.html}{https://cxc.cfa.harvard.edu/csc/columns/spectral\_properties.html}} for sources with $>$150 counts in the 0.5$-$7\,keV energy range. In these cases the powerlaw models generally provide a superior fit over the blackbody models based on the reduced statistic. Therefore, we assume powerlaw components ($S$\,$\sim$\,$E^{-\Gamma}$) to describe the accreting sources. 

In our models we attenuate these powerlaw components by photoelectric absorption in the Milky Way Galaxy along the line of sight ($n_{\rm{H,Gal}}$; estimated from the \colden~function within \ciao) and by absorption intrinsic to the host galaxy ($n_{\rm{H,exgal}}$). If the \CSCTwo~provides a powerlaw spectral index ($\Gamma_{\mathrm{CSC}}$) and a column density ($n_{\mathrm{H,CSC}}$) for a source, then we set $\Gamma=\Gamma_{\mathrm{CSC}}$ and $n_{\rm{H,exgal}}=n_{\mathrm{H,CSC}}-n_{\rm{H,Gal}}$. Otherwise, we fix the spectral index at $\Gamma$\,$=$\,2.1 and the intrinsic absorption to $n_{\rm{H,exgal}}$\,$=$\,$3\times10^{21}$\,\uNH~ \citep[see e.g.][and references therein]{Walton:2022}. Unabsorbed, rest-frame X-ray luminosities (\LXBUnabs) are computed using the host galaxy redshifts (Section \ref{sec:galselect}) and cosmology stated in Section \ref{sec:intro}.

To select X-ray sources that are consistent with the traditional definition of ULXs, we impose a lower luminosity limit of \LXBUnabs\,$=$\,$10^{39}$\,\uLum~(conventional lower threshold for ULX selection corresponding to the approximate theoretical Eddington limit for a 10\,\Msun~BH). To avoid AGN and likely IMBH candidates \citep[e.g. ESO 243-49 HLX-1;][]{Farrell:2009}, we also impose an upper X-ray luminosity limit of $2\times10^{41}$\,\uLum. This limit is motivated by the currently most luminous known ULX that is confirmed to be associated with accretion onto a stellar mass object (\citealp{Israel:2017}; converted to \LXBUnabs~from the peak 0.3$-$10\,keV luminosity assuming a powerlaw spectrum with a photon index of $\Gamma$\,$=$\,$2.1$). As in \citet{Walton:2011} and \citet{Earnshaw:2019}, to retain the largest sample of ULX candidates we also include sources that are consistent with these lower and upper luminosity thresholds when accounting for their upper and lower uncertainties, respectively. These limits yield \ULXASZ~spatially offset X-ray sources that satisfy the luminosity criteria for ULXs.

\begin{deluxetable*}{lcccccc}
\tabletypesize{\footnotesize}
\tablecolumns{7}
\tablecaption{ULX Candidates.}
\tablehead{
\colhead{\CSCTwo~Source} &
\colhead{\sdss~Host Galaxy} &
\colhead{\z} &
\colhead{Offset} &
\colhead{$L_{\mathrm{0.5-7\,keV}}$} &
\colhead{\Mstar} &
\colhead{\SFR} \\
\colhead{($-$)} &
\colhead{($-$)} &
\colhead{($-$)} &
\colhead{(kpc)} &
\colhead{($10^{39}$\,\uLum)} &
\colhead{(log[\Mstar/\Msun])} &
\colhead{(log[\SFR/\uSFR])} \\
\colhead{1} &
\colhead{2} &
\colhead{3} &
\colhead{4} &
\colhead{5} &
\colhead{6} &
\colhead{7}
}
\startdata
2CXO J000120.2$+$130641 & J000119.98$+$130640.59 & $0.018^a$ & $1.39\pm0.20$ & $4.9_{-1.5}^{+1.5}$ & $12.0\pm0.41$ & $-3.6\pm3.80$ \\
2CXO J000131.2$+$233409 & J000131.33$+$233403.97 & $0.070^a$ & $7.99\pm1.94$ & $98.0_{-30.3}^{+30.3}$ & $9.7\pm1.17$ & $0.3\pm0.56$ \\
2CXO J000846.5$+$192147 & J000846.76$+$192146.84 & $0.138^a$ & $6.95\pm1.58$ & $144.0_{-123.4}^{+123.5}$ & $9.7\pm1.09$ & $0.1\pm0.37$ \\
2CXO J001335.6$-$192804 & J001335.55$-$192805.75 & $0.148^a$ & $7.76\pm2.05$ & $26.8_{-23.9}^{+22.9}$ & $11.2\pm0.09$ & $1.5\pm0.01$ \\
2CXO J002231.2$+$002110 & J002231.07$+$002109.59 & $0.04715^b$ & $2.24\pm0.71$ & $6.0_{-2.6}^{+2.6}$ & $8.8\pm0.03$ & $-2.7\pm0.01$ \\
2CXO J003413.6$-$212803 & J003414.03$-$212811.00 & $0.02329^b$ & $4.51\pm0.40$ & $6.7_{-3.5}^{+3.3}$ & $10.5\pm0.09$ & $0.8\pm0.01$ \\
2CXO J004852.6$+$315735 & J004852.84$+$315731.08 & $0.017^a$ & $1.68\pm0.21$ & $0.7_{-0.3}^{+0.3}$ & $9.8\pm0.38$ & $1.5\pm0.08$ \\
2CXO J004947.9$+$321632 & J004947.81$+$321639.80 & $0.01555^b$ & $2.21\pm0.28$ & $1.7_{-0.5}^{+0.5}$ & $11.5\pm0.48$ & $1.0\pm0.48$ \\
2CXO J005513.9$+$352600 & J005513.99$+$352603.00 & $0.03683^b$ & $2.27\pm0.46$ & $4.5_{-2.4}^{+2.4}$ & $10.4\pm0.16$ & $1.1\pm0.60$ \\
2CXO J011505.5$+$002546 & J011505.90$+$002546.84 & $0.030^a$ & $2.88\pm0.82$ & $16.2_{-7.7}^{+7.7}$ & $8.5\pm0.22$ & $0.0\pm0.09$
\enddata
\tablecomments{Column 1: ULX candidate \CSCTwo~source; column 2: ULX candidate host galaxy; column 3: best available redshift of the host galaxy in column 2; column 4: ULX candidate projected physical offset from the host galaxy centroid; column 5: ULX candidate unabsorbed, rest-frame 0.5$-$7\,keV luminosity; columns 6\,$-$\,7: ULX candidate host galaxy stellar mass (\Mstar) and star formation rate (\SFR).\\ $^a$Photometric redshift\\ $^b$Spectroscopic redshift\\ (This table is available in its entirety in machine-readable form.)}
\label{tab:ULXCat}
\end{deluxetable*}

\subsection{Removing AGN and Known Contaminants}
\label{sec:AGN}

We remove ULX candidates that are likely to be AGN (i.e. accreting massive BHs) based on their mid-infrared (MIR) colors by crossmatching them with the \wisetitle~\citep[\wise;][]{Wright:2010} using a radius equal to five times the X-ray source positional uncertainty. After applying the 90\%~completeness criterion defined in \citet{Assef:2018}, we then identify and remove eight MIR AGN \citep[this filter may miss some low luminosity AGN; e.g.][]{Hickox:2009,Barrows:2021}. We also remove four X-ray sources that have stellar counterparts in the \pnstrstitle~(\pnstrs) \iband-band imaging \citep[determined by applying models composed of Sersic components and a background following the procedure in][]{Barrows:2019} since they may actually be associated with a distinct galaxy and hence are likely AGN. Any possible remaining AGN will have large X-ray-to-optical ratios or be in remnant stellar cores that have undergone significant tidal stripping (this scenario is more likely for IMBHs; see Section \ref{sec:intro}).

We also crossmatch the ULX candidates with \ned~(using the same crossmatch radius of five times the X-ray source positional uncertainty) to search for any association with known extended radio jets or gravitationally-lensed AGN since they can mimic offset X-ray sources; none are found. Using the same \ned~crossmatch we identify and remove one ULX candidate that is coincident with a source possessing a spectroscopic redshift (\z\,$=0.53$) that is significantly different from that of its candidate host galaxy (\z\,$=0.01$).

As noted in \citet{Earnshaw:2019} and \citet{Walton:2022}, host galaxy AGN may contaminate ULX samples if they are not coincident with the optical centroid of the host galaxy. Therefore, we follow the approach taken in those works to quantify the spatial offsets of X-ray AGN. Of the subset of matches between X-ray sources and galaxies (Section \ref{sec:initmatch}), we identify X-ray AGN as sources with unabsorbed, rest-frame 2$-$10 keV luminosities ($L_{\mathrm{2-10\,keV}}$) of $L_{\mathrm{2-10\,keV}}\ge10^{42}$\,\uLum~\citep[converted from the observed 0.5$-$7\,keV flux using the same procedure as described in Section \ref{sec:xrayspec}, assuming a powerlaw spectrum and a typical AGN photon index of $\Gamma$\,$=$\,1.7; e.g.][]{Middleton:2008}. We then determine the relative uncertainties on the X-ray source offsets (as in Section \ref{sec:regastr}). The X-ray AGN offsets are found to be smaller than the 3$\sigma$ uncertainties in all cases, indicating that our statistical threshold is sufficient to exclude nuclear AGN.

Finally, as noted in \citet{Walton:2022}, off-axis observations can bias the \CSCTwo~fluxes toward high values due to large PSFs (and hence large extraction radii) that may incorporate emission from other sources in crowded fields or extended emission. Therefore, we visually inspect the extraction radii of each observation that contributes to the aperture flux for a Master Source (i.e. all OBSIDs in the ‘best’ Bayesian block) and remove any sources for which the extraction regions incorporate clearly unrelated sources or extended emission. One source is flagged and removed in this step.

\subsection{Final Sample}
\label{sec:final}

The final sample contains \ULXSZ~unique ULX candidates among \GalSZ~host galaxies. The ULX candidates, host galaxies, redshifts, projected physical offsets, and intrinsic 0.5$-$7\,keV luminosities are listed in Table \ref{tab:ULXCat}. While ULX candidates hosted by low-redshift galaxies have been extensively cataloged \citep{Colbert:2002,Swartz:2004,Liu:Bregman:2005,Liu:2005,Liu.J:2011,Swartz:2011,Gong:2016,Earnshaw:2019,Kovlakas:2020,Inoue:2021,Walton:2022}, few are known out to intermediate redshifts. This sample increases the known number of ULX candidates past \z\,$\sim$\,0.15 by a factor of $\sim$\,3 and significantly past \z\,$\sim$\,0.3 for the first time. 

The redshift distribution is shown along the top axis of Figure \ref{fig:LX_DELTA_S_Z_PLOT} and spans the range \z\,$=$\,\ZMin\,$-$\,\ZMax , where \SpecPerc\%~(\SpecSZ) have spectroscopic redshifts, and the remaining \PhotPerc\%~(\PhotSZ) have only photometric redshifts. The majority of spectroscopic redshifts obtained from the literature (i.e. not from the \sdss) are for relatively nearby galaxies, leading to the observed bias of spectroscopic redshifts toward lower values compared to photometric redshifts. The low redshift limit of the sample reflects the distribution of nearby \sdss~galaxies. Due to the \CSCTwo~sensitivity limits, a redshift-dependent luminosity bias exists (upper plot of Figure \ref{fig:LX_DELTA_S_Z_PLOT}). Moreover, due to the angular resolution limits imposed by the \ch~PSF (Section \ref{sec:xrayselect}) and the relative astrometric accuracy (Section \ref{sec:regastr}), a redshift-dependent physical offset bias also exists (lower plot of Figure \ref{fig:LX_DELTA_S_Z_PLOT}).

Studies of nearby ULXs often reveal several candidates within a host galaxy \citep[i.e. a mean of $\sim$\,2;][]{Ptak:2004,Swartz:2004,Liu:2005,Swartz:2011,Earnshaw:2019,Walton:2022}. When limiting our sample to a similar volume (\z\,$<$\,0.05), the mean number of ULX candidates per host galaxy (\NULXPerHostMean) is lower by comparison (Figure \ref{fig:NULX_PER_GAL_PLOT}), potentially due to systematic differences in the measured sizes of host galaxy extents, to different criteria used for the X-ray source selection, or a combination thereof. Figure \ref{fig:overlays_multiple} shows examples of host galaxies with multiple ULX candidates. More distant hosts typically contain only one identified ULX candidate due to X-ray imaging sensitivity limits and to the PSFs that limit the selection of off-nuclear sources \citep{Hornschemeier:2004,Lehmer:2006,Mainieri:2010}. We similarly observe a declining number of ULX candidates per host galaxy with increasing redshift (Figure \ref{fig:NULX_PER_GAL_PLOT}). Examples of host galaxies with only one ULX candidate, extending out to the maximum redshift of our sample, are shown in Figure \ref{fig:overlays}.

Additional quality flags are available for Master Sources in the \CSCTwo~(\texttt{likelihood\_class}, \texttt{sat\_src\_flag}, \texttt{dither\_warning\_flag}, \texttt{streak\_src\_flag}, and \texttt{pileup\_flag}), some of which have been incorporated by previous catalogs of nearby ULX candidates from the \CSCTwo~\citep{Kovlakas:2020,Walton:2022}. Among our final sample, \MARGSZ~(\MARGPERC\%) have a \texttt{likelihood\_class} value of \texttt{MARGINAL}, while the remaining \TRUESZ~(\TRUEPERC\%) have a value of \texttt{TRUE}. Only one source has a \texttt{dither\_warning\_flag} set, and none of the other flags are set. Since detections near the flux sensitivity limits will more often be \texttt{MARGINAL}, we include both types in our sample to probe higher redshifts. Our subsequent qualitative conclusions regarding contamination fractions (Section \ref{sec:cont}), host galaxy properties (Section \ref{sec:sfr_mstar}), comparison with XRBs (Section \ref{sec:com_xrb}), and occupation fractions (Section \ref{sec:frac}) remain unchanged when the sample is limited to sources with \texttt{likelihood\_class}\,$=$\,\texttt{TRUE} and with no other flags. However, in Section \ref{sec:cxb} we discuss the impact of these flags on our results regarding the ULX contribution to the CXB.

\begin{figure}[t!]
\includegraphics[width=0.475\textwidth]{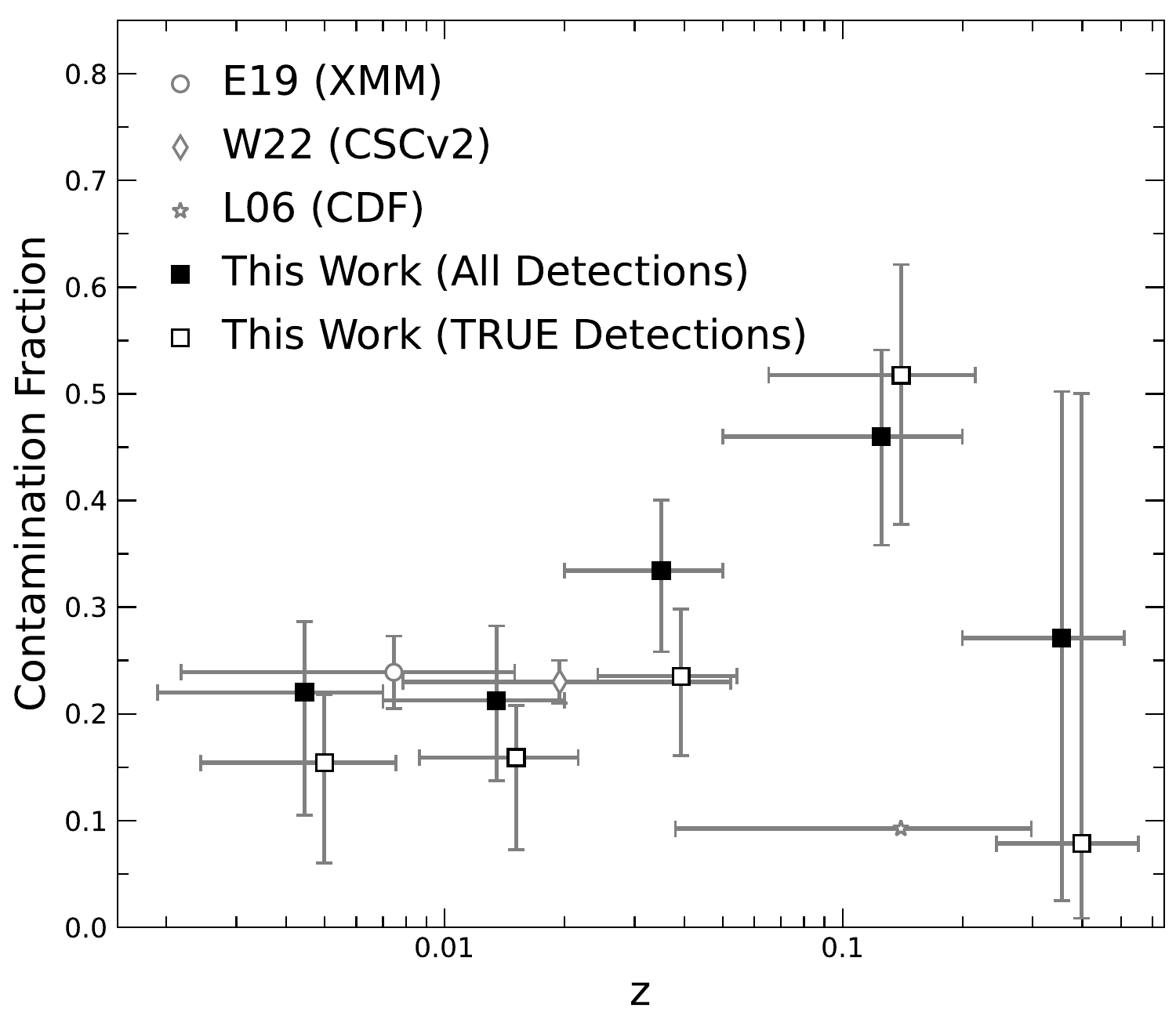}
\vspace{-0.2in}
\caption{\footnotesize{Contamination fraction against redshift (\z) for our full sample of ULX candidates (filled squares) and the subset with \texttt{likelihood\_class} values of TRUE and without any quality flags set (open squares, horizontally offset for clarity). The redshift bins are defined in Figure \ref{fig:NULX_PER_GAL_PLOT}. Vertical errorbars represent the upper and lower 68.3\%~binomial confidence intervals, and horizontal errorbars represent the bin width. We also show comparison estimates from a \ch~Deep Field sample (\citealp{Lehmer:2006}; L06), an XMM-Newton sample (\citealp{Earnshaw:2019}; E19), and a \CSCTwo~sample (\citealp{Walton:2022}; W22), plotted at the sample mean or median distances and with uncertainties shown (if published). When accounting for uncertainties, our estimates are consistent with those from previous catalogs of both nearby and intermediate-redshift ULX candidates.}}
\label{fig:cont_frac_z}
\end{figure}

\subsection{Diffuse Emission and Source Confusion}
\label{sec:confusion}

We compute the 0.5$-$7\,keV emission expected from hot ISM gas using the host galaxy \SFRs~(Section \ref{sec:sfr_mstar}), the 0.3$-$10\,keV relation from \citet{Mineo:2012}, and a thermal powerlaw index of $\Gamma$\,$=$\,3 \citep[e.g.][]{Mezcua:2016,Barrows:2019}. We then scale this value by the ratio of the \ch~PSF area to the total galaxy area (subtended by the Petrosian radius). In Sections \ref{sec:XRB_SFR_Mstar} and \ref{sec:cxb} we remove this estimated hot gas contribution when examining how the ULX candidate luminosities and fluxes evolve with redshift.

A single \CSCTwo~detection may represent multiple physically distinct X-ray sources that are unresolved by \ch, particularly at higher redshifts. To test this effect on our results, for each ULX candidate we compute the value of \LXBUnabs~that would be detected assuming the host galaxy has the same X-ray point source population as the Antennae galaxy system \citep[a merger between NGC 4038 and NGC 4039 and a prototypical XRB/ULX-rich system due to the recent merger-triggered star formation; e.g.][]{Fabbiano:2001,Zezas:2006}. We use the list of Antennae X-ray sources from \citet{Poutanen:2013} and convert the 0.1$-$10\,keV~luminosities to \LXBUnabs~using the best-fit powerlaw spectral indices from \citealp{Zezas:2002}. For each ULX candidate, we determine the total emission from X-ray sources in the Antennae (placed at the host galaxy redshift) that would be confused due to the \CSCTwo~PSF ($L_\mathrm{{0.5-7,Conf}}$). To account for spatial variations of the Antennae X-ray point source population, we compute $L_\mathrm{{0.5-7,Conf}}$ centered at 1,000 random positions within the Antennae light profile \citep[as defined in][]{Poutanen:2013}. 

The maximum number of confused X-ray point sources from these estimates reaches 11 at the highest redshifts of our sample due to the significant physical extents of the \ch~PSF profiles. This suggests that the luminosities of some of our ULX candidates may be due to the integrated emission of several ULXs plus diffuse emission from hot ISM gas. In Section \ref{sec:frac} we use these results to correct the ULX occupation fraction estimates for source confusion.

\begin{figure}[t!]
\includegraphics[width=0.48\textwidth]{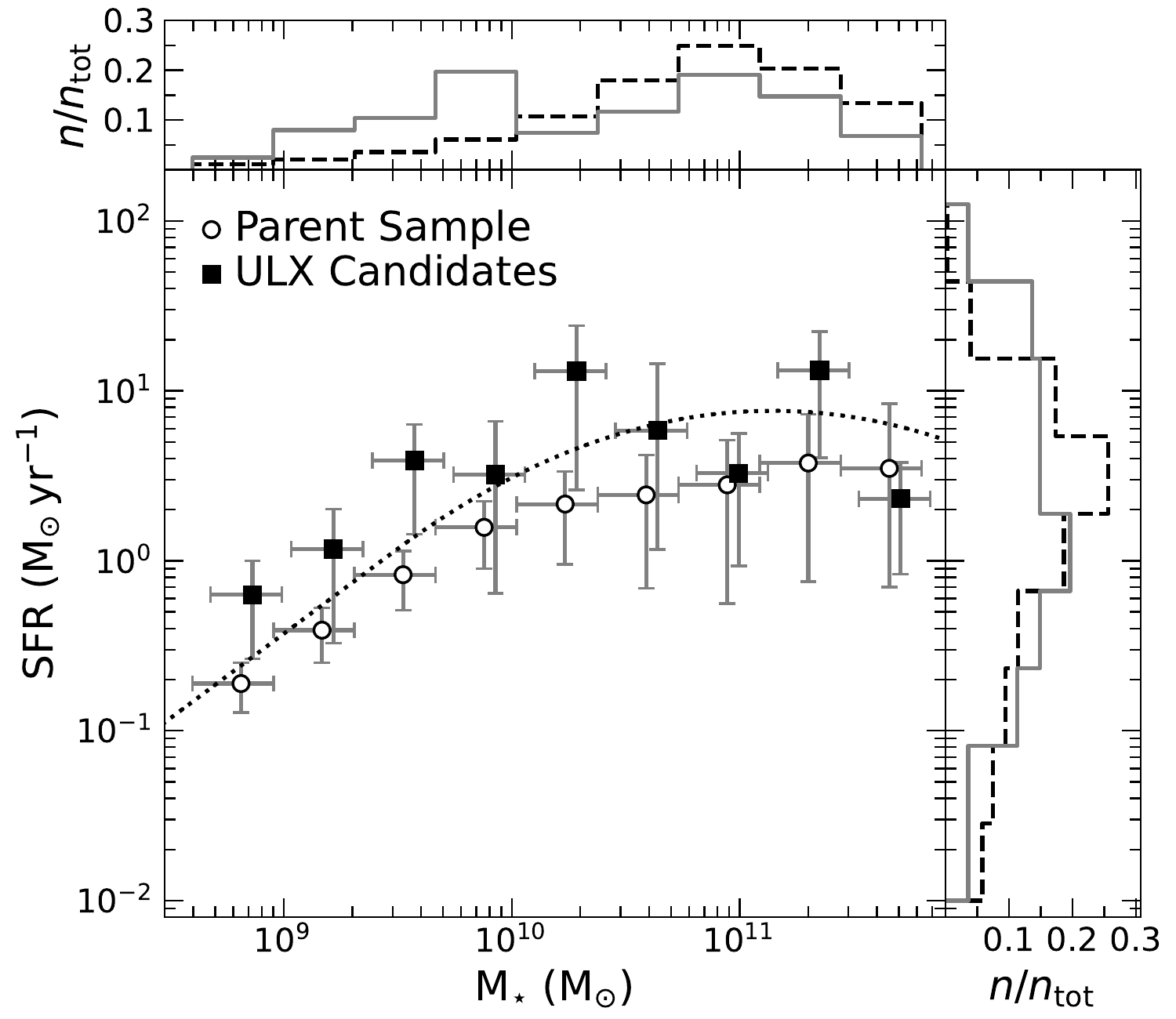}
\vspace{-0.2in}
\caption{\footnotesize{Galaxy star formation rate (\SFR) against stellar mass (\Mstar). The dotted line shows the redshift-dependent relation from \citet{Schreiber:2015}, computed using the median redshift of the parent galaxy sample. The data points represent median \SFRs~of the parent galaxy sample (open circles) and ULX candidate hosts (filled squares) in bins of even logarithmic spacing along the abscissa (horizontally offset for clarity). The vertical errorbars represent the two-sided standard deviation within the bin, and the horizontal errorbars denote the bin width. Histograms for values along the ordinate and abscissa (each normalized to unity) are shown on the right and top, respectively, for the parent sample (black dashed) and ULX candidate hosts (gray solid). Both samples are generally consistent with the relation for star forming galaxies, though the specific \SFRs~of ULX candidate hosts are systematically elevated relative to those of the parent sample.}}
\label{fig:SFR_MSTAR_PLOT}
\end{figure}

\begin{figure}[t!]
\includegraphics[width=0.475\textwidth]{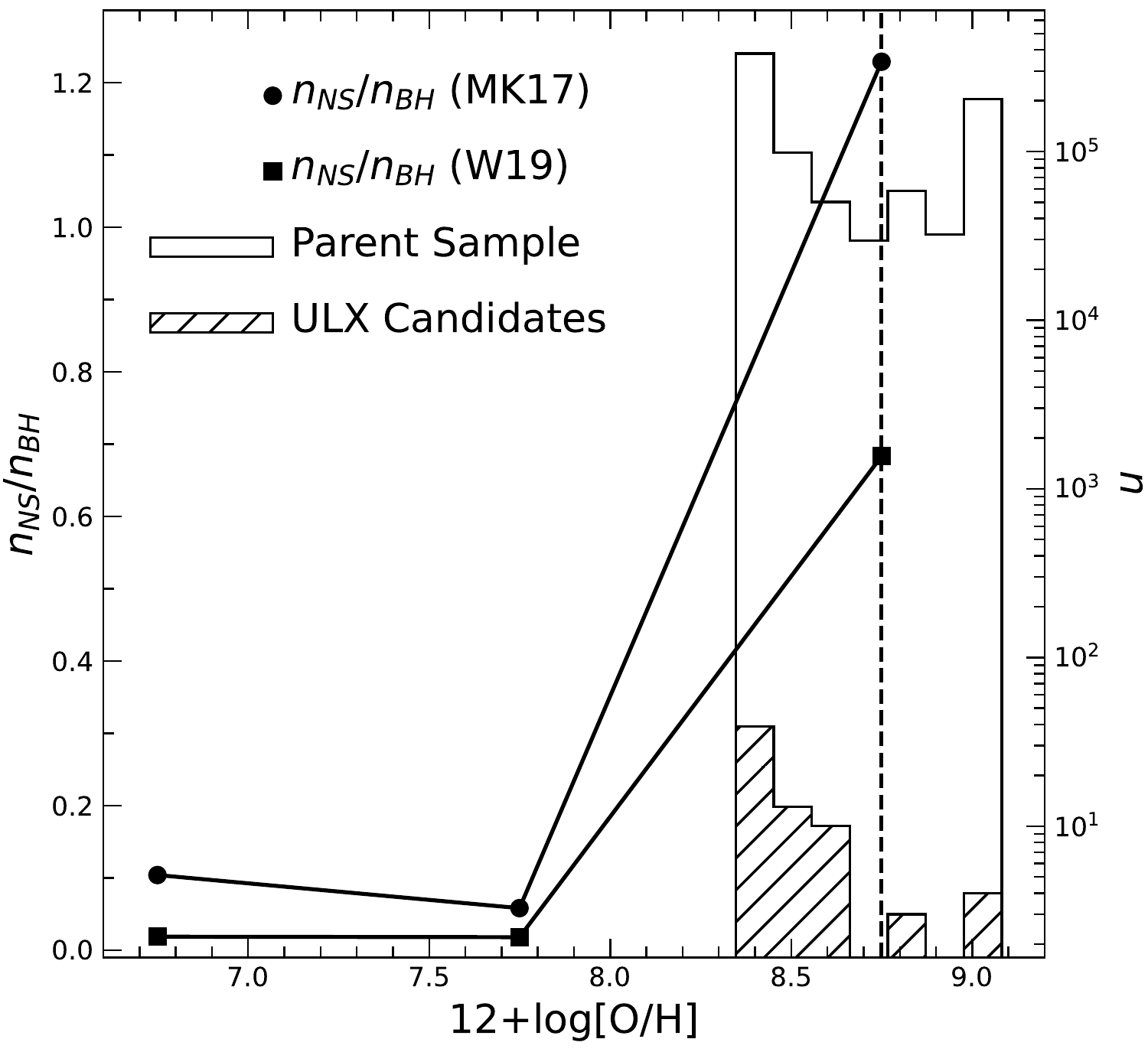}
\vspace{-0.2in}
\caption{\footnotesize{Left ordinate: predicted ratio of neutron star to BH acretors ($n_{NS}/n_{BH}$) as a function of host galaxy metallicity (O/H) from the population synthesis models of \citet{Middleton:2017} (MK17; filled circles) and \citet{Wiktorowicz:2019} (W19; filled squares). Right ordinate: number ($n$) of galaxies in our parent sample (open histogram) and in the sample of ULX candidate hosts (hatched histogram). Metallicity estimates are derived from fitted stellar population models (Section \ref{sec:met}) that consist of seven different values: $\mathrm{log}[O/H]+12=8.35, 8.45, 8.55, 8.65, 8.75, 8.85, \mathrm{and}\,8.95$. The vertical dashed line indicates Solar metallicity. The ULX host galaxies show a preference for low metallicities relative to the parent sample, and they are mostly consistent with sub-Solar metallicities. The predicted $n_{NS}/n_{BH}$ ratios suggest at least some contribution from BH accretors.}}
\label{fig:MET_PLOT}
\end{figure}

\section{Contamination Fractions}
\label{sec:cont}

While known contaminants are removed in Section \ref{sec:AGN}, here we estimate the number of unknown background or foreground sources that remain. Following the methodology applied to previous nearby ULX catalogs \citep{Walton:2011,Sutton:2012,Earnshaw:2019,Walton:2022}, we compute the number of X-ray sources expected to randomly be within the area of each galaxy in the parent sample (a circle defined by the Petrosian radius) minus the inner offset threshold rectangle (defined by three times the offset uncertainty in right ascension and declination). The parent galaxy offset uncertainties are the quadrature sum of the relative astrometric uncertainty between the \sdss~image and the overlapping \ch~images (computed as described in Section \ref{sec:regastr}) and the average \CSCTwo~source centroid error from our final sample.

We determine the expected number of X-ray sources using the resolved 0.5$-$7\,keV point source density function of \citet{Masini:2020} and the effective limiting sensitivities at each galaxy position. We set the effective limiting sensitivity to the flux corresponding to an observed 0.5$-$7\,keV luminosity of $10^{39}\,$erg s$^{-1}$ at the host galaxy redshift, or otherwise the 0.5$-$7\,keV \CSCTwo~limiting sensitivity at the galaxy position if it is larger (where the \CSCTwo~limiting sensitivities are obtained from the \CSCTwo~all-sky limiting sensitivity map that corresponds to the deepest sensitivity value among all stacks that cover a given position\footnote{\href{https://cxc.cfa.harvard.edu/csc/char.html}{https://cxc.cfa.harvard.edu/csc/char.html}}). We omit any parent galaxies with \CSCTwo~limiting sensitivities corresponding to greater than the maximum luminosity of our selection (\LXBUnabs\,$=2\times10^{41}$\,\uLum). The contamination fraction is then the total number of expected X-ray sources in the parent sample divided by the total number of ULX candidates (with the AGN identified in Section \ref{sec:AGN} removed from both quantities).
 
Figure \ref{fig:cont_frac_z} shows how the expected fraction of contaminants varies with redshift. The fractions for the full sample and the subset with \texttt{likelihood\_class} values of TRUE and without any quality flags set (see Section \ref{sec:final}) are consistent when accounting for the uncertainties. Moreover, our estimates are statistically consistent with those from previous samples at both low and intermediate redshifts. The overall mean contamination fraction is \ContFracwErrsb\%~and no statistically significant evidence for redshift evolution is detected.

\begin{figure}[t!]
\includegraphics[width=0.475\textwidth]{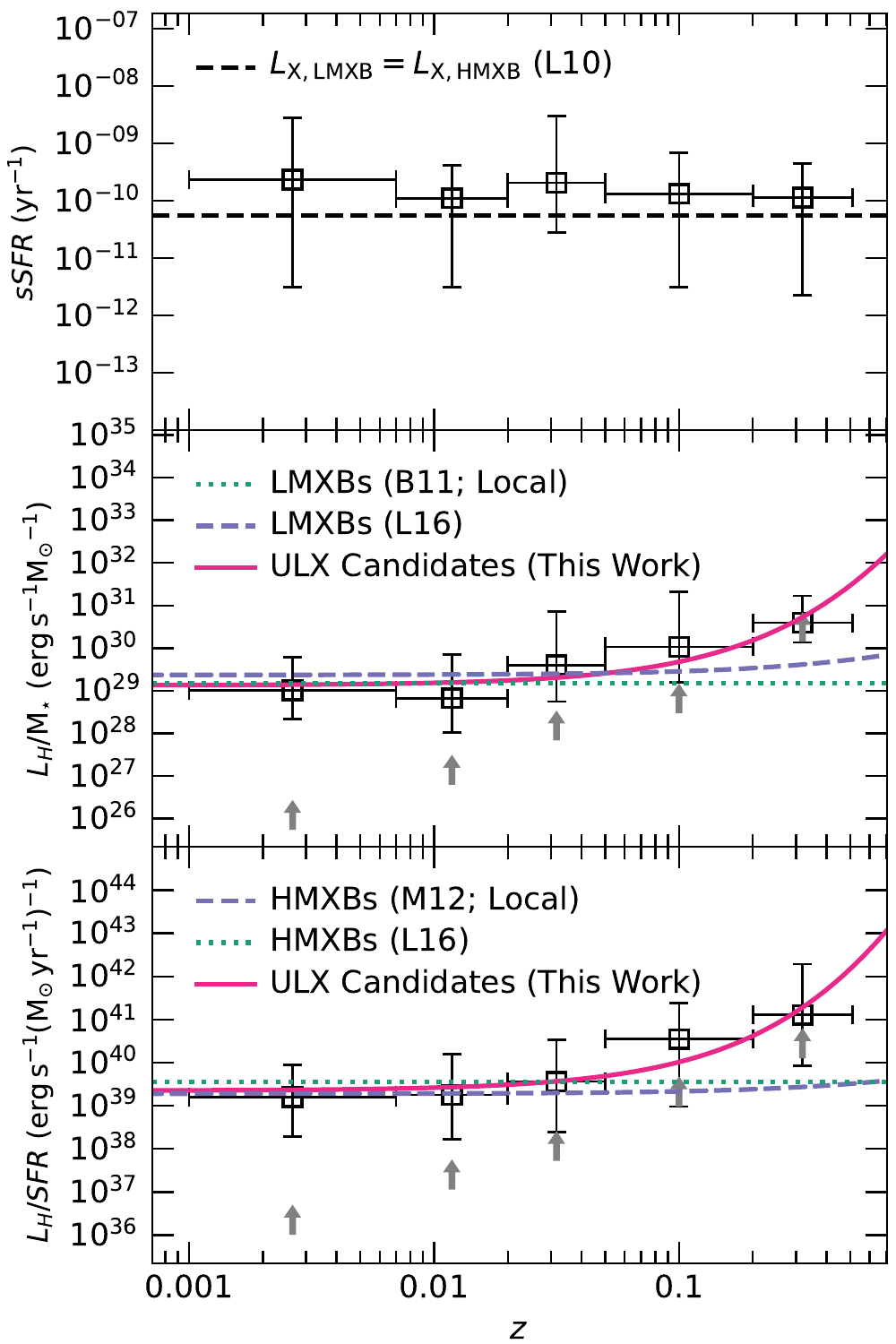}
\vspace{-0.2in}
\caption{\footnotesize{Top: Specific \SFR~(\sSFR) against redshift. The black dashed line denotes the value of \sSFR~above which HMXBs dominate the X-ray emission of their host galaxies, relative to LMXBs, from \citet{Lehmer:2010} (L10). At each redshift, the \sSFRs~are consistent with similar contributions though systematically deviate toward a larger HMXB contribution. Middle and bottom panels: hard X-ray (2$-$10) luminosity over host galaxy stellar mass ($L_{H}$/\Mstar) and \SFR~($L_{H}$/\SFR). The green dotted lines represent the local relations from \citet{Boroson:2011} (B11; middle) and \citet{Mineo:2012} (M12; bottom), the purple dashed lines represent the redshift-dependent relations from \citet{Lehmer:2016} (L16), and the magenta solid lines represent the best-fit powerlaw functions to our sample. In all panels, the open black squares represent median values from our sample within the same redshift bins defined in Figure \ref{fig:NULX_PER_GAL_PLOT}, the vertical errorbars denote the standard deviation within the bins, and the horizontal errorbars denote the bin widths. The gray arrows denote average lower limits for our sample in each bin. Due to the luminosity limits, the scaling relations for our ULX candidate sample evolve more quickly with redshift (compared to XRBs).}}
\label{fig:LXoSFR_SSFR_Z_BIN_PLOT}
\end{figure}

\section{Host Galaxy Properties}
\label{sec:sfr_mstar}

Stellar masses (\Mstar) and \SFRs~for each host galaxy are computed by applying models to broadband spectral energy distributions (SEDs) using the Code Investigating GALaxy Emission \citep[\cigale;][]{Noll:2009,Boquien:2019} that accounts for absorbed and re-radiated starlight through an energy balance approach. To build the SEDs, we supplement the \sdss~photometry with detections from the \galextitle~\citep[\galex;][]{Bianchi1999}, the \twomasstitle~\citep[\twomass;][]{Skrutskie:2006}, and \wise~(using a matching radius of $2''$). Our models assume a delayed star formation history, a Salpeter initial mass function \citep{Salpeter:1955}, and the stellar population libraries of \citet{Bruzual:Charlot:2003}. Lower and upper bounds on the best-fit parameters are the 16th and 84th percentiles determined from fits to synthetic data created from random Gaussian distributions with standard deviations equal to the photometry uncertainties. 

We run additional models that also include an AGN component, and an AGN is considered to be present if the F$-$distribution probability is greater than 90\%. This corresponds to six galaxies, none of which show evidence for a nuclear X-ray source. Hence, their ULX candidates may instead be massive BHs with AGN-like SEDs. Alternatively, a heavily-obscured AGN (undetected in X-rays) may be present, or an AGN component is simply inaccurately included in the SED model due to poor photometry or fitting results.

The values of \Mstar~and \SFR~for the ULX candidate host galaxies are listed in Table \ref{tab:ULXCat}. \SFR~is plotted against \Mstar~in Figure \ref{fig:SFR_MSTAR_PLOT}, and (when accounting for uncertainties) both the parent and ULX candidate hosts are in agreement with expectations from normal star-forming galaxies. The ULX candidate hosts show a systematic offset toward larger specific \SFRs~(\sSFRs), where the median offset is $\Delta \sSFR$\,$=$\,\MedDeltasSFR\,dex. While the statistical significance of the offset is weak in each stellar mass bin, the trend is consistent with observed preferences for ULXs to be found in galaxies with relatively high \SFRs~\citep[e.g.][]{Swartz:2009}.

\begin{figure}[t!]
\includegraphics[width=0.475\textwidth]{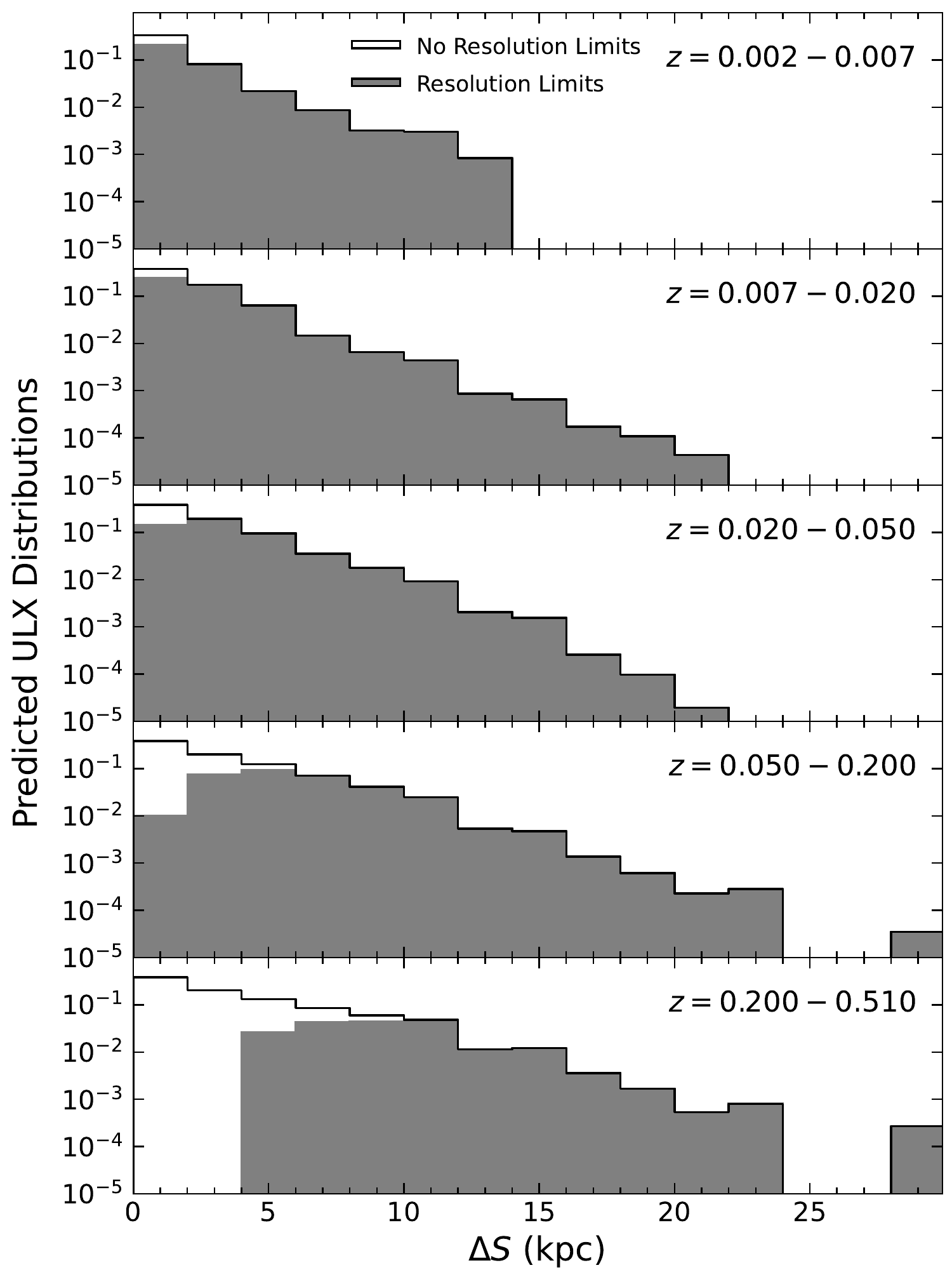}
\vspace{-0.2in}
\caption{\footnotesize{Predicted number of ULXs in our sample (normalized to a sum of unity) that could be found in our sample, with (solid gray) and without (open) the resolution limits imposed, as a function of projected physical separation (\DeltaS) and in the same redshift bins defined in Figure \ref{fig:NULX_PER_GAL_PLOT}. The fraction of recovered ULXs is the ratio of the two distributions, and it declines with increasing redshift. The larger physical offsets observed at higher redshifts are a result of the bias illustrated in the lower panel of Figure \ref{fig:LX_DELTA_S_Z_PLOT}.}}
\label{fig:OFFSET_RES_PLOT_Z}
\end{figure}

\begin{figure}
\includegraphics[width=0.475\textwidth]{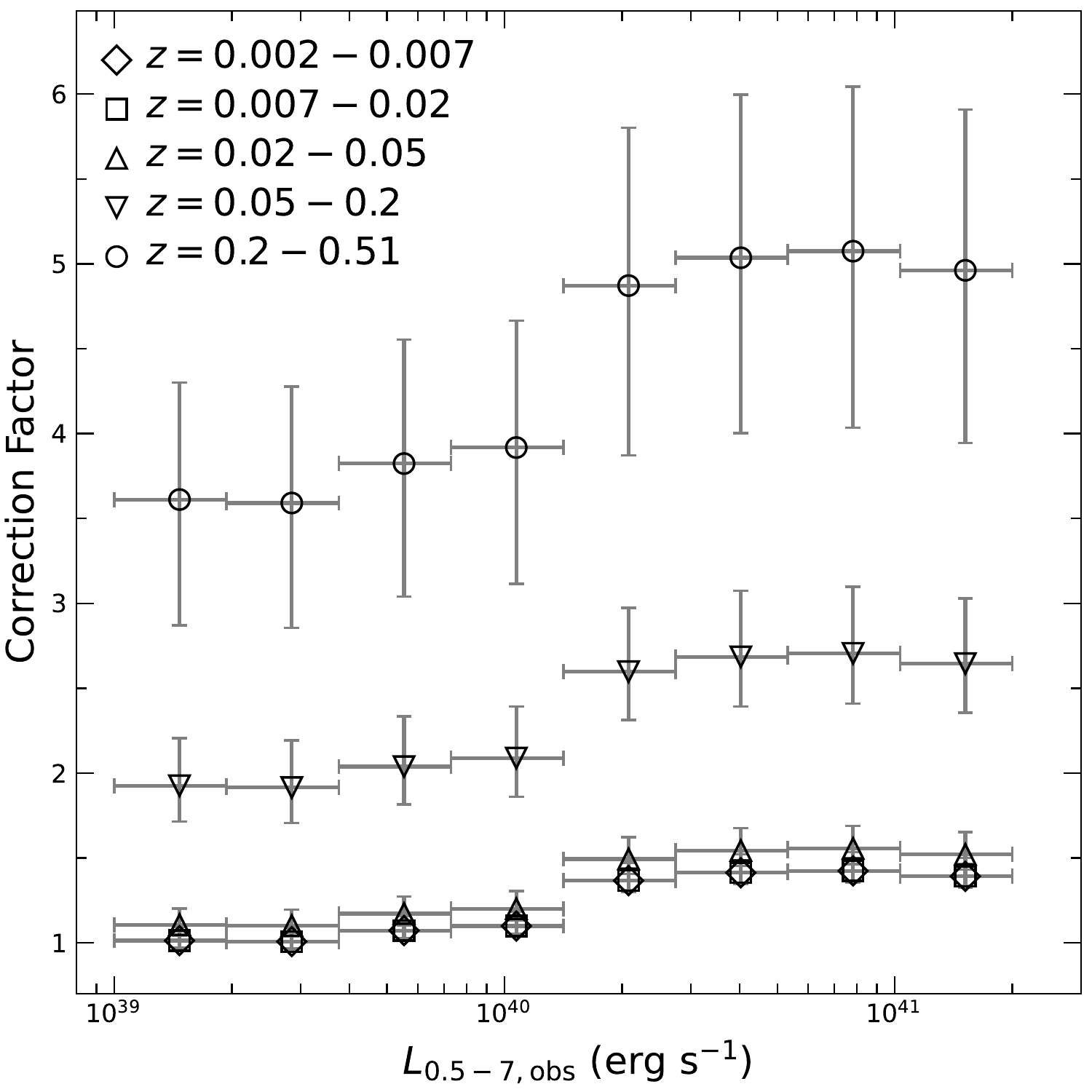}
\vspace{-0.2in}
\caption{\footnotesize{Spatial resolution correction factor for the observed ULX occupation fractions, computed as a function of observed 0.5$-$7\,keV luminosity ($L_{\mathrm{0.5-7,obs}}$) and in the five redshift bins defined in Figure \ref{fig:NULX_PER_GAL_PLOT}. The vertical errorbars denote the 68.3\%~binomial confidence intervals, and the horizontal errorbars denote the bin widths.}}
\label{fig:SCALE_FAC_PLOT}
\end{figure}

\section{Comparison to XRB Populations}
\label{sec:com_xrb}

Several studies of local ULXs have identified their optical counterparts to be massive stars \citep[e.g.][]{Motch:2014,Heida:2015,Heida:2019b}, suggesting that ULXs may be preferentially associated with HMXBs. Indeed, the results from Section \ref{sec:sfr_mstar} and Figure \ref{fig:SFR_MSTAR_PLOT} show that the ULX candidate hosts exhibit a systematic offset toward large \sSFRs~(relative to the parent galaxy sample) that is expected if HMXBs dominate the sample. However, as suggested by the presence of ULXs in early-type galaxies \citep[e.g.][]{Plotkin:2014}, in some cases they may instead be associated with LMXBs. 

Since the spatial resolution of our host galaxy imaging precludes the identification of stellar counterparts, to constrain the nature of our ULX candidates we instead compare their global host galaxy properties with those of XRBs. In Section \ref{sec:met} we examine their host galaxy metallicities, and in Section \ref{sec:XRB_SFR_Mstar} we determine how their X-ray powers scale with their host galaxy stellar masses and \SFRs.

\subsection{Host Galaxy Metallicities}
\label{sec:met}

Decreasing metallicities are observed to correlate with increasing numbers of ULXs per host galaxy \citep[e.g.][]{Zampieri:2009,Mapelli:2010,Prestwic:2013,Douna:2015,Kovlakas:2020}, consistent with a significant contribution from HMXBs. Theoretical results also predict that metallicity plays a strong role in the type of accretors found in ULXs, with larger neutron star (versus BH) contributions for higher metallicities \citep[e.g.][]{Middleton:2017,Wiktorowicz:2019}.

\begin{figure*}[t!]
\hspace*{0.08in} \includegraphics[width=0.97\textwidth]{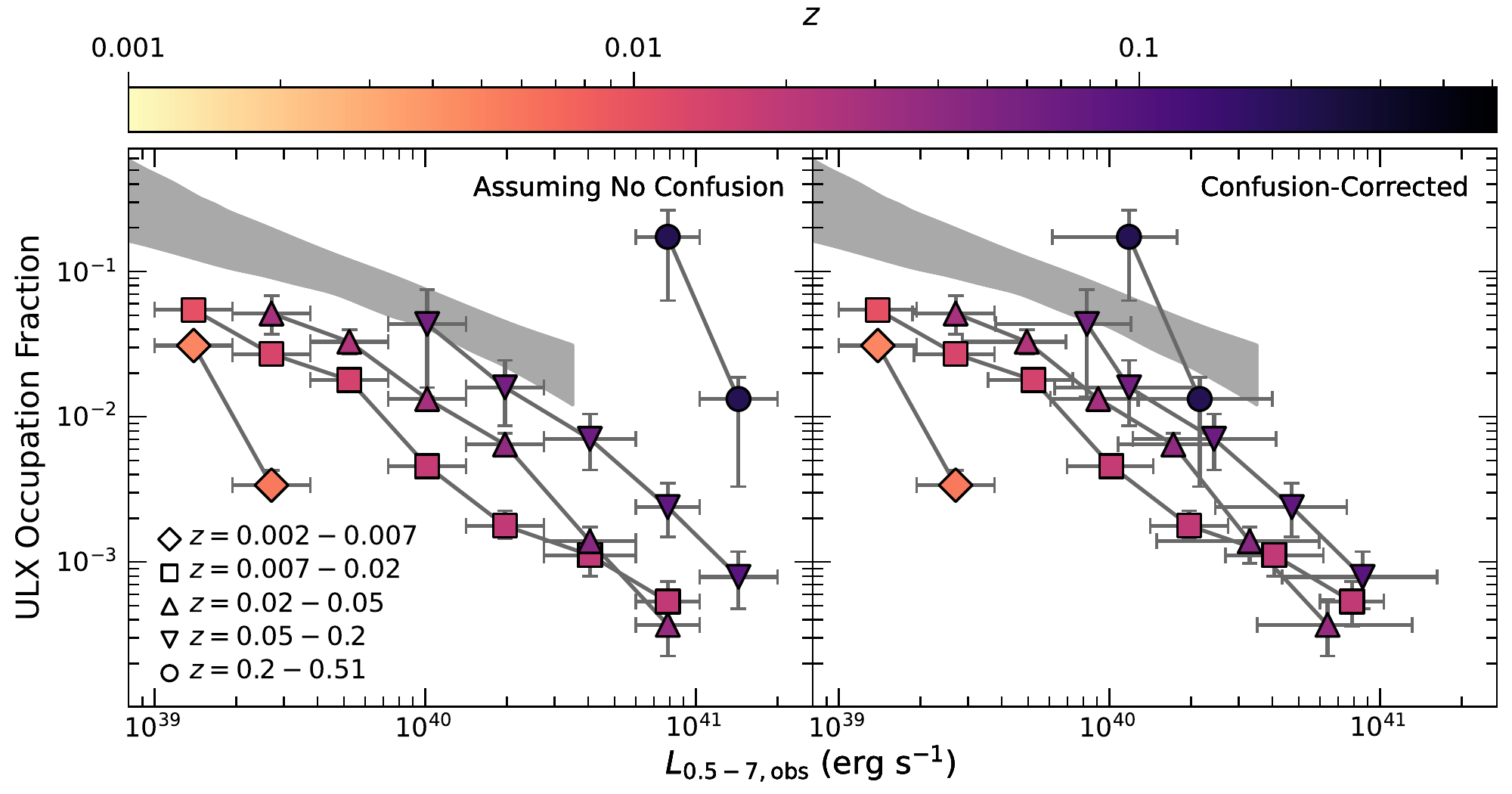}
\vspace{-0.05in}
\caption{\footnotesize{ULX occupation fraction against observed 0.5$-$7\,keV luminosity ($L_{\mathrm{0.5-7,obs}}$) in the five redshift bins defined in Figure \ref{fig:NULX_PER_GAL_PLOT}. These fractions are obtained by correcting the observed ULX occupation fraction for spatial resolution (Figure \ref{fig:SCALE_FAC_PLOT}), contamination (Section \ref{sec:cont}), and galaxy inclinations (Section \ref{sec:initmatch}). The left and right panels, respectively, show fractions without and with corrections for the median estimated number of confused sources applied (see Section \ref{sec:frac} for details). The vertical errorbars denote the 68.3\%~confidence intervals (quadrature sum of the binomial bounds, the contamination fraction uncertainties, and the correction factor uncertainties), and the horizontal errorbars denote the bin widths (plus the minimum and maximum bounds on the confusion estimates in the right panel). The data points are color-coded according to the median redshift of the ULX candidates in each bin. The gray-shaded region shows the 1$\sigma$ bounds on the corrected ULX occupation fraction from the intermediate-redshift sample (\z\,$=$\,0.038\,$-$\,0.298) of \citet{Lehmer:2006}.
}}
\label{fig:ULX_FRAC_PLOT}
\end{figure*}

For the subset of ULX candidate hosts in our sample that were spectroscopically-selected as galaxies in the Baryon Oscillation Spectroscopic Survey \citep[BOSS;][]{Dawson:2013}, metallicity estimates are available from model grids \citep[generated using Flexible Stellar Population Synthesis;][]{Conroy:2009} fitted to the \sdss~photometry \citep{Montero-Dorta:2016}. When using the subset of these model results that allow for an extended star-formation history and account for dust, the metallicities of our ULX candidate host galaxies show a bias toward low values relative to the parent sample (a two-sample KS-test yields a null hypothesis probability of $10^{-13}$ that they are identical) that suggests a preference for hosting HMXBs (Figure \ref{fig:MET_PLOT}). This is also consistent with the observed over-abundances of ULX candidates in low-mass galaxies \citep[and hence low-metallicity environments; e.g.][]{Griffith:2011,Remy-Ruyer:2015,O'Connor:2016} reported in previous studies \citep[e.g.][]{Swartz:2008,Kovlakas:2020} and apparent in the top histogram of Figure \ref{fig:SFR_MSTAR_PLOT} for our sample. 

The host galaxy metallicities are predominantly sub-Solar, consistent with samples of nearby HMXBs \citep[e.g.][]{Brorby:2016,Fornasini:2019,Fornasini:2020,Lehmer:2021}. When compared to predictions from the population synthesis models of \citet{Middleton:2017} and \citet{Wiktorowicz:2019}, our results are generally consistent with contributions from both neutron star and BH accretors. These predictions vary negligibly over the redshift range of our sample. \\

\subsection{Scaling Relations with Host Galaxy Stellar Mass and \SFR}
\label{sec:XRB_SFR_Mstar}

Since LMXB and HMXB X-ray emissivity scale with host galaxy \Mstar~and \SFR, respectively, the \sSFR~provides a strong indicator of their relative X-ray contributions. The evolution of \sSFR~with \z~is shown in the top panel of Figure \ref{fig:LXoSFR_SSFR_Z_BIN_PLOT}, and the values are statistically consistent with similar contributions from both LMXBs and HMXBs \citep[\sSFR\,$=$\,5.6\,$\times$\,$10^{-11}$\,\usSFR;][]{Lehmer:2010} over the full sample redshift range. However, a systematic offset toward a stronger HMXB contribution is observed (median value of \sSFR\,$=$\,1.3\,$\times$\,$10^{-10}$\,\usSFR). 

The middle and bottom panels of Figure \ref{fig:LXoSFR_SSFR_Z_BIN_PLOT} show how our sample compares to the hard X-ray (2$-$10\,keV) LMXB and HMXB galaxy scaling relations ($L_{H}$/\Mstar~and $L_{H}$/\SFR, respectively) for local samples \citep{Boroson:2011,Mineo:2012} and as a function of redshift \citep{Lehmer:2016}: $L_{H}$/\Mstar\,$=$\,$\alpha$\,(1\,+\,\z)$^{\gamma}$ and $L_{H}$/\SFR\,$=$\,$\beta$\,(1\,+\,z)$^{\delta}$ (where $\alpha$\,$=$\,$10^{29.37\pm0.15}$\,\ualpha, $\beta$\,$=$\,$10^{39.28\pm0.05}$ \ubeta, $\gamma$\,$=$\,$2.03\pm0.60$, and $\delta$\,$=$\,$1.31\pm0.13$). Hard X-ray luminosities for our sample are computed using the same spectral models described in Section \ref{sec:xrayspec}, and they are corrected for the contribution from hot ISM gas (Section \ref{sec:confusion}). When parameterized with the same powerlaw functional form, our sample yields consistent normalizations of $\alpha$\,$=$\,\LXMstaralpha\,\ualpha~and $\beta$\,$=$\,\LXSFRbeta\,\ubeta~(offsets at the \alphaSigOff$\sigma$ and \betaSigOff$\sigma$ levels, respectively). This agreement suggests that, for nearby host galaxies, ULX luminosities have a similar dependence on host stellar mass and \SFR~as do typical XRB populations. 

However, the fits to our sample yield significantly larger powerlaw slopes of $\gamma$\,$=$\LXMstargamma~and $\delta$\,$=$\LXSFRdelta~(offsets at the \gammaSigOff$\sigma$ and \deltaSigOff$\sigma$ levels, respectively). As indicated by the lower limits shown in the middle and bottom panels of Figure \ref{fig:LXoSFR_SSFR_Z_BIN_PLOT}, this stronger redshift evolution is likely reflecting that our sample is biased toward luminous X-ray sources at higher redshifts (e.g. the top panel of Figure \ref{fig:LX_DELTA_S_Z_PLOT}) compared to the deeper stacks from \citet{Lehmer:2016}. Therefore, at higher redshifts our procedure is selecting a subset of luminous LMXBs and HMXBs with highly efficient accretion rates. While the larger ratios of X-ray luminosity to stellar mass and \SFR~may possibly be elevated due to AGN contamination, an AGN component is not favored in the SED models (Section \ref{sec:sfr_mstar}) of any host galaxies above \z\,$=$\,0.15 (where our sample significantly deviates from that of XRB populations). Moreover, any such AGN would not be MIR-detected, nor would they have optical counterparts detected by the \pnstrs~imaging (Section \ref{sec:AGN}). Hence, if AGN they are likely powered by accretion onto IMBHs.

\section{Redshift Evolution of the ULX Occupation Fraction}
\label{sec:frac}

For galaxies in the nearby Universe, \citet{Hornschemeier:2004} estimate that the ULX occupation fraction (fraction of galaxies that host at least one ULX candidate) is $8^{+8}_{-5}\%$, and \citet{Ptak:2004} find a similar value of up to $\sim$\,10\,$-$\,20\%. On the other hand, at higher redshifts (\z\,$=$\,0.03\,$-$\,0.25) \citet{Hornschemeier:2004} estimate a fraction of $36^{+24}_{-15}\%$ in the \ch~Deep Fields. Over a similar redshift range and with an augmented sample, \citet{Lehmer:2006} estimate a fraction of up to $\sim$\,30\%. Moreover, when compared to a matched local sample from \citet{Ptak:2004}, they find the intermediate-redshift ULX occupation fraction to be elevated at the 80\%~level. If ULXs are predominantly associated with HMXBs, their increasing galaxy occupation fractions with redshift may be explained by the increase in \SFR~comoving density with redshift due to higher cold gas fractions, lower metallicities, and merger-triggered star-formation \citep[e.g.][and references therein]{Madau:2014}. Enhanced galaxy merger rates may also lead to the presence of off-nuclear accreting IMBHs with faint stellar cores that would contribute to the observed ULX fraction.

However, these results are based on comparisons between multiple different samples that introduce a heterogeneous set of selection biases. For instance, using an intermediate-redshift sample of ULX candidates from the COSMOS field, \citet{Mainieri:2010} find a ULX occupation fraction that is several times to an order of magnitude lower than that from the \ch~Deep Field samples, potentially due to shallower flux limits that inhibit detection of the lowest luminosity ULXs. Therefore, we use our sample of uniformly-selected ULXs over the range \z\,$=$\,\ZMin\,$-$\,\ZMax~to estimate the redshift evolution of the ULX occupation fraction in a consistent manner and out to \z\,$\sim0.5$ for the first time.

Following the procedures from \citet{Ptak:2004} and \citet{Lehmer:2006}, in logarithmically-spaced luminosity bins over the range $10^{39}$\,$-$\,$2\times10^{41}$\,\uLum, we first determine the number of parent sample galaxies with limiting 0.5$-$7\,keV luminosities (see Section \ref{sec:cont}) greater than or equal to the bin lower edge. In each bin the ULX occupation fraction is the number of those galaxies with a ULX of observed 0.5$-$7\,keV luminosity equal to or greater than the bin lower limit divided by the total number of galaxies (we are using observed luminosity, rather than unabsorbed, rest-frame luminosity, since that is directly comparable to the limiting luminosity).

\begin{figure}[t!]
\includegraphics[width=0.475\textwidth]{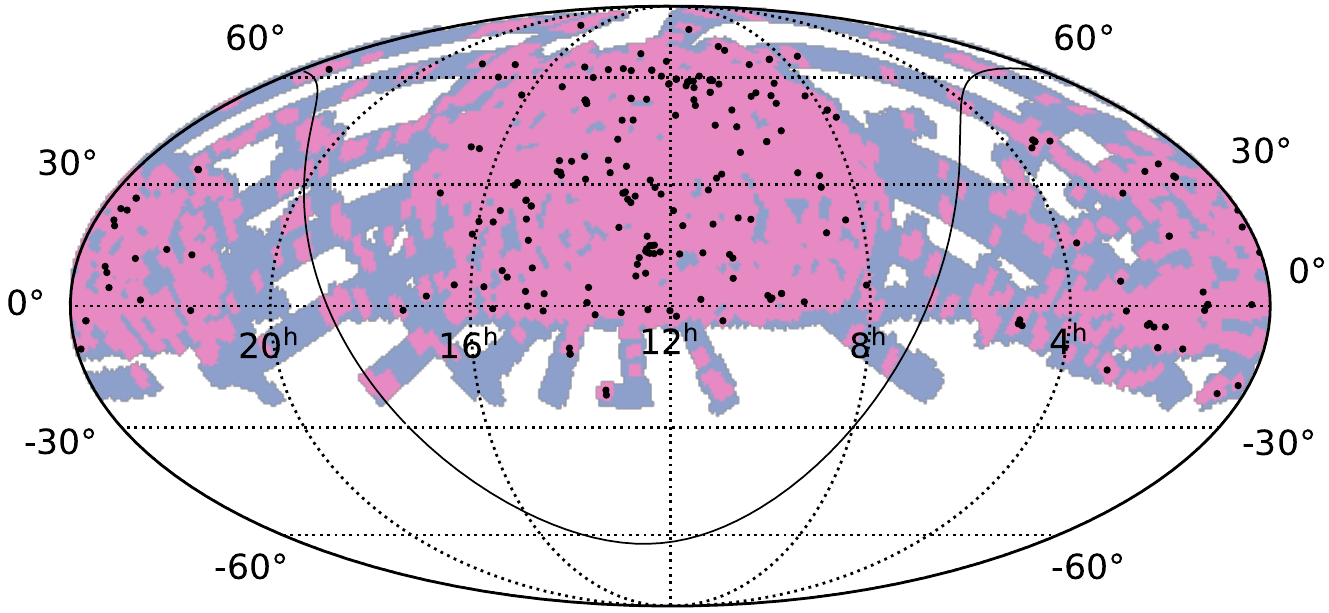}
\vspace{-0.2in}
\caption{\footnotesize{Sky map of the \sdss~DR16 (gray), the overlapping coverage from the \CSCTwo~(pink), and our individual ULX candidates (black). The map is in equatorial coordinates and shown with a Mollweide projection. The Galactic Plane is indicated by the solid line.
}}
\label{fig:SKY_MAP_CSC}
\end{figure}

\begin{figure*}
\includegraphics[width=\textwidth]{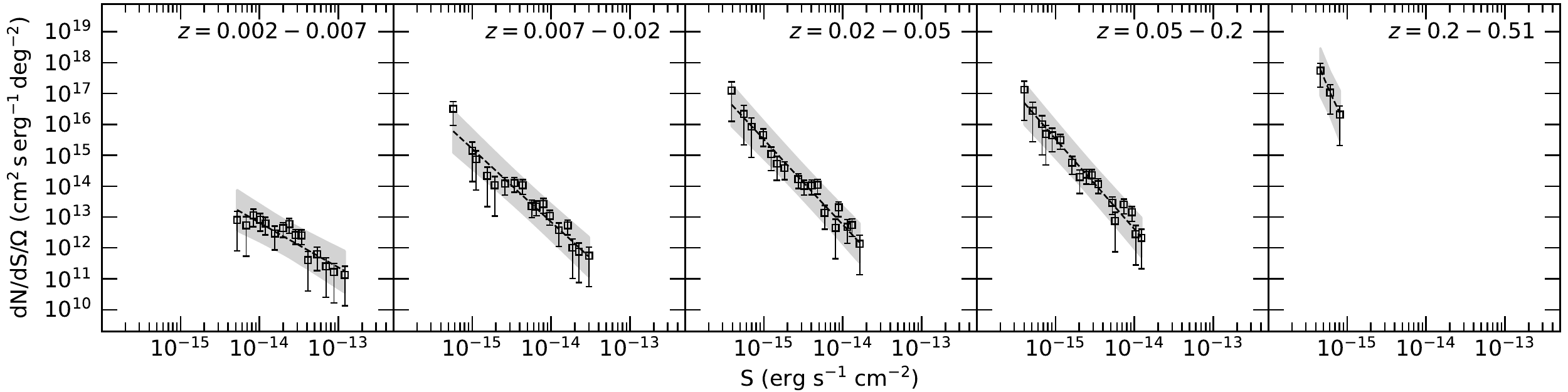}
\vspace{-0.2in}
\caption{\footnotesize{Normalized differential source counts ($dN/dS/\Omega$) as a function of flux ($S$) in the redshift bins defined in Figure \ref{fig:NULX_PER_GAL_PLOT}. The counts have been corrected for spatial resolution (Figure \ref{fig:SCALE_FAC_PLOT}, averaged over $L_{\mathrm{0.5-7,obs}}$\,$=$\,$10^{39}$\,$-$\,$2\times10^{41}$\,\uLum), contamination (Section \ref{sec:cont}), and galaxy inclination (Section \ref{sec:initmatch}). The black dashed line is the best-fit powerlaw function, and the shaded region represents the 3$\sigma$ two-sided confidence bounds.}}
\label{fig:dN_dS_OMEGA}
\end{figure*}

The \ch~angular resolution introduces a bias that misses ULXs with small projected physical offsets at high redshifts (i.e. the lower panel of Figure \ref{fig:LX_DELTA_S_Z_PLOT}). To correct for this bias, we follow the procedure from \citet{Lehmer:2006} that consists of estimating the number of our ULX candidate host galaxies for which an offset X-ray source could be detected, both with and without imposing the resolution limits (i.e. the offset uncertainties) as a function of projected physical separation. We then convolve both of these distributions with the normalized true distribution of ULXs (taken from \citealt{Kovlakas:2020} and restricted to a distance of $<$\,40\,Mpc and nuclear offsets of $>\,$0.05\,kpc for a complete sample not limited by spatial resolution) to obtain the predicted true distributions for our sample. Since the resolution limits have a greater impact on the detectable physical offsets at higher redshifts, we compute these distributions as function of redshift (Figure \ref{fig:OFFSET_RES_PLOT_Z}). The ratio of these distributions yields the spatial resolution correction factor as a function of redshift. As in \citet{Lehmer:2006}, we also correct for the different numbers of ULXs per galaxy between our sample and that of \citet{Kovlakas:2020} as a function of observed luminosity. The final spatial resolution correction factors are shown in Figure \ref{fig:SCALE_FAC_PLOT}.

After correcting for spatial resolution limits, in addition to contamination (Section \ref{sec:cont}) and galaxy inclination (Section \ref{sec:initmatch}), the ULX occupation fractions are shown in the left panel of Figure \ref{fig:ULX_FRAC_PLOT}. Our estimated fractions decrease with increasing luminosity, consistent with the negative correlations observed both locally \citep{Ptak:2004} and at intermediate redshifts \citep{Lehmer:2006,Mainieri:2010}. The redshift-dependent luminosity bias imposed by the \CSCTwo~sensitivity limits (i.e. the top panel of Figure \ref{fig:LX_DELTA_S_Z_PLOT}) is reflected in the trend of increasing median ULX candidate redshift with luminosity and in the absence of lower luminosity sources at higher redshifts. 

Source confusion is discussed in Section \ref{sec:confusion} using the Antennae merging galaxy system as a template, and here we use those results to correct for this effect. We first subtract the expected hot ISM gas contribution from each ULX candidate. Assuming the remaining emission is a superposition of luminosities from multiple X-ray sources, and that this number is equal to the median number of confused Antennae X-ray point sources, we divide it by this number to obtain the corrected luminosities of the sources contributing to the observed emission (assuming the observed emission is distributed equally among the confused sources). Then, the luminosities of each bin are adjusted based on the mean correction factor in each bin. The corrected luminosities of the occupation fractions are shown in the right panel of Figure \ref{fig:ULX_FRAC_PLOT}, with the corrections being strongest at higher redshifts. The horizontal errorsbars include the range of values obtained when assuming the minimum and maximum estimated numbers of confused sources to convey the uncertainties associated with these corrections.

For a given X-ray luminosity, the ULX occupation fractions increase with redshift, even after correcting for confusion. This trend is qualitatively consistent with that inferred from comparisons of local and intermediate-redshift samples \citep{Hornschemeier:2004,Lehmer:2006,Mainieri:2010} despite the different samples and X-ray energy ranges used. Moreover, our sample shows that this trend continues until at least \z\,$\sim$\,0.5. The confusion-corrected results suggest that the number of galaxies hosting at least one ULX of a given luminosity increases by a factor of $\sim$\,2 from \z\,$\sim$\,0.3 to \z\,$\sim$\,0.5, roughly consistent with the corresponding increase in the \SFR~comoving density over that redshift interval \citep[e.g.][]{Hopkins:2006,Madau:2014}. In Section \ref{sec:cxb} we discuss the implications of this evolution on the relation between ULX luminosities and host galaxy properties. The possibility of accretion onto IMBHs in some cases also remains (though, as stated in Section \ref{sec:XRB_SFR_Mstar}, this is not supported by the host galaxy SED models).

\begin{figure}
\includegraphics[width=0.475\textwidth]{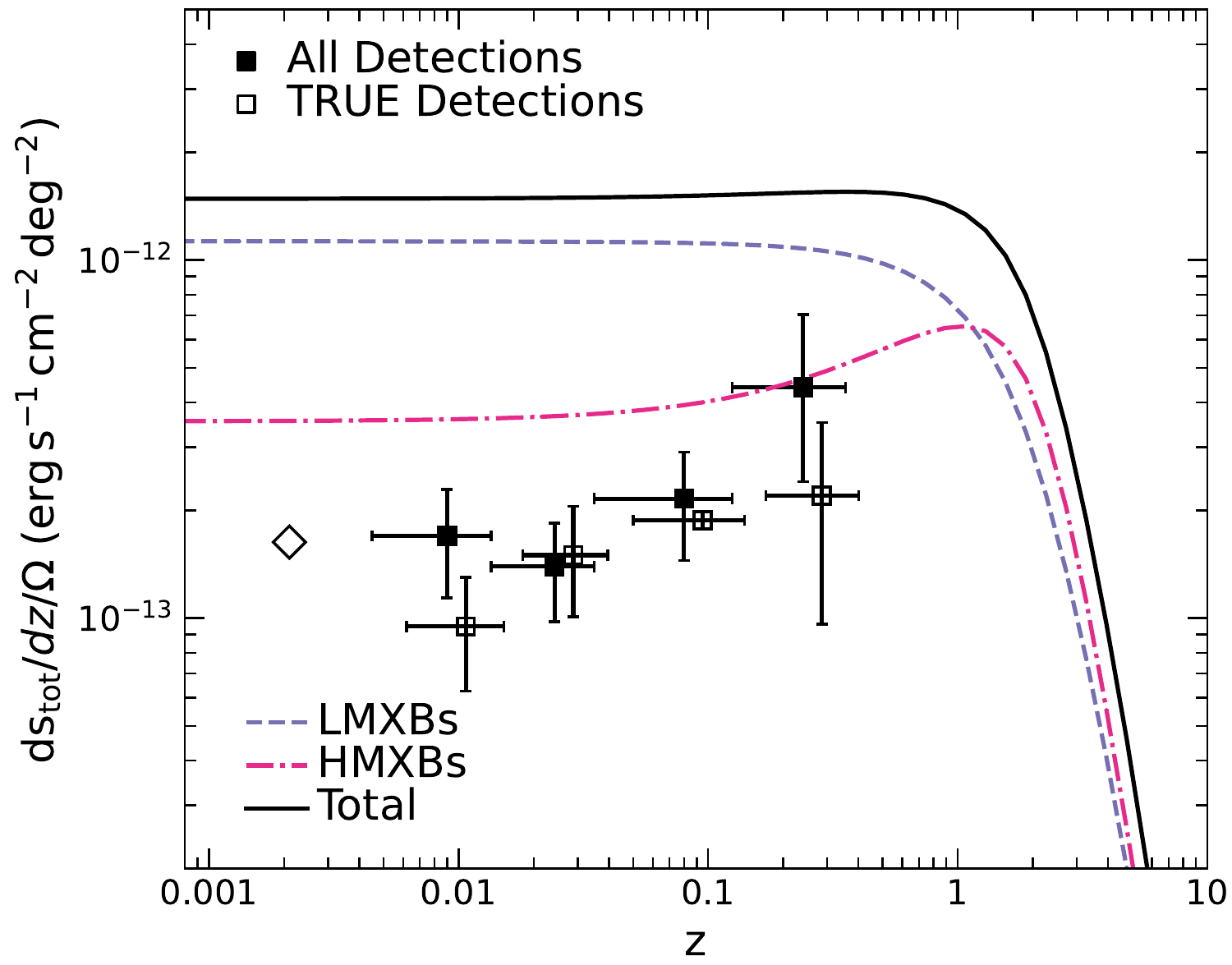}
\vspace{-0.2in}
\caption{\footnotesize{Change in normalized flux per unit redshift ($ds_{\mathrm{tot}}/dz/\Omega$) as a function of redshift (\z) for our full sample of ULX candidates (filled squares) and the subset with \texttt{likelihood\_class} values of TRUE and without any quality flags set (open squares, horizontally offset for clarity). The vertical errorbars indicate the 68.3\%~confidence intervals, and the horizontal errorbars denote the total redshift range of sources contributing to the differential. The individual LMXB and HMXB X-ray point source contributions \citep{Lehmer:2016} (L16; from detections and stacking) are shown as blue dashed and magenta dot-dashed lines, respectively, and their sum is shown by the black solid line. For reference, the value from the complete sample of \citet{Swartz:2011} (S11; converted to flux per unit solid angle using their sample median redshift) is also shown (open daimond). The X-ray background contribution from our sample of ULXs is consistent with local measurements. At \z\,$\sim$\,0.5, ULXs account for up to $\sim$\,40\%~of the total galaxy X-ray background flux.}}
\label{fig:XRAY_Z_PLOT}
\end{figure}

\begin{figure}
\includegraphics[width=0.475\textwidth]{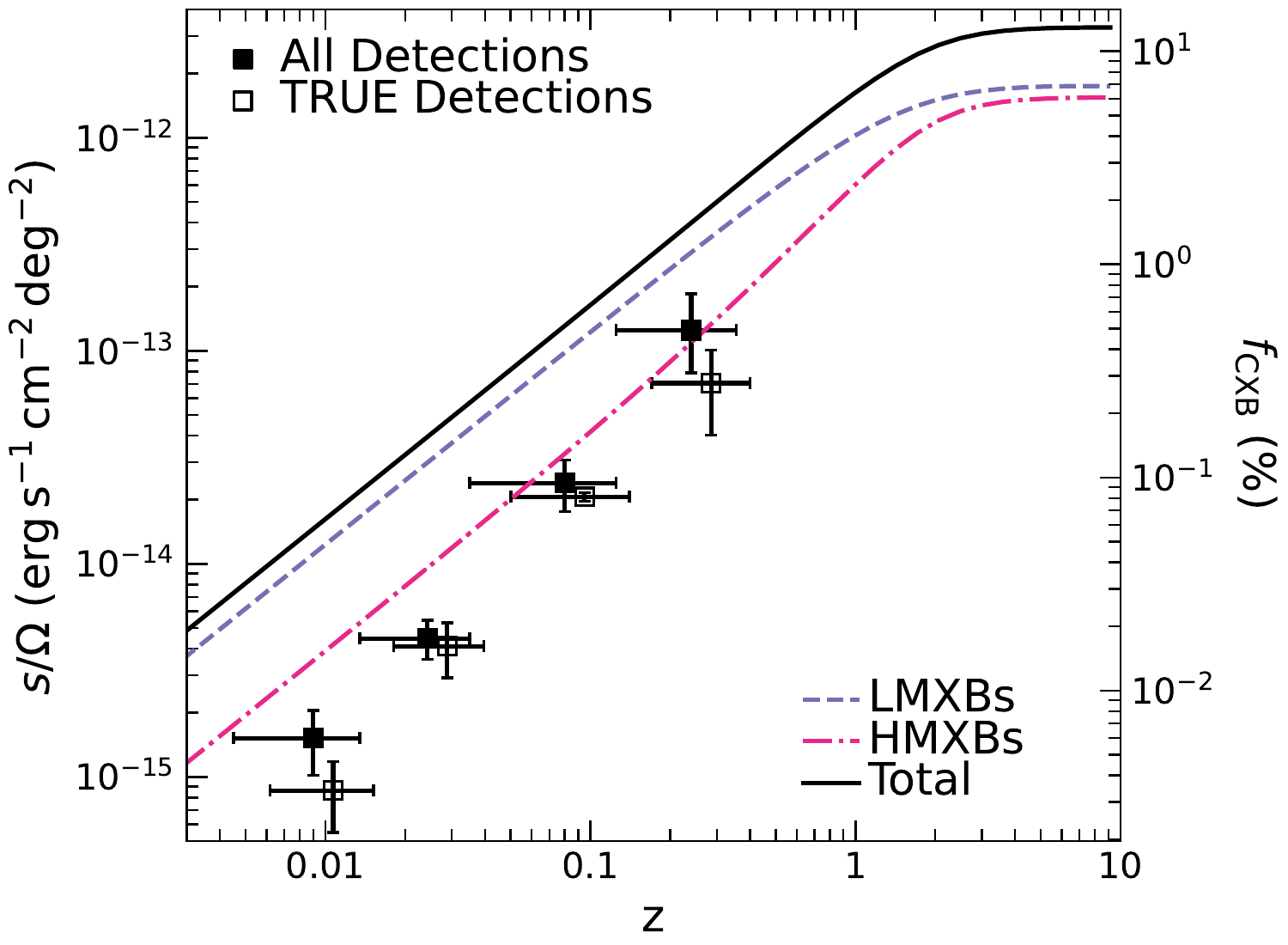}
\vspace{-0.2in}
\caption{\footnotesize{Cumulative sum of the normalized 0.5$-$7\,keV flux ($s/\Omega$), computed from the differential values shown in Figure \ref{fig:XRAY_Z_PLOT}. The equivalent percent contribution to the CXB ($f_{\mathrm{CXB}}$) is shown on the right ordinate, where the CXB value is the summed soft (0.5\,$-$\,2\,keV) and hard (2\,$-$\,8\,keV) intensities derived in \citet{Lehmer:2012}. The vertical errorbars indicate the 68.3\%~confidence intervals (quadrature sum of the binomial bounds, the contamination fraction uncertainties, and the correction factor uncertainties), and the horizontal errorbars denote the bin widths. The marker styles and reference curves are the same as in Figure \ref{fig:XRAY_Z_PLOT}. Out to \z\,$=$\,0.51, the integrated ULX contribution to the CXB is $\sim$\,1\%, and is expected to be $\sim$\,5\%~if it follows that of star-forming galaxies.}
}
\label{fig:CXB_Z_PLOT}
\end{figure}

\section{ULX Contribution to the CXB}
\label{sec:cxb}

While the majority of the CXB is from AGN emission \citep[for a review see][]{Brandt:2005}, the normal galaxy contribution is estimated to be up to $\sim$\,20\%~\citep[depending on the energy range and assumptions regarding the X-ray emissivity spectral shape; e.g.][]{Natarajan:2000,Hickox:2006,Lehmer:2012,Dijkstra:2012,Aird:2015,Lehmer:2016}. In normal galaxies without significant star formation, the X-ray emission is dominated by LMXBs, and at low redshifts this represents a significant fraction of the total X-ray emission from the normal galaxy population. Toward increasingly higher redshifts, HMXBs begin to dominate the normal galaxy X-ray emission due to the increasing comoving \SFR~density \citep[peaking at $z\sim$\,2; e.g.][]{Hopkins:2006,Madau:2014}. If ULXs are posited to trace LMXBs and HMXBs, they will likewise contribute significantly to the ionizing radiation output of normal galaxies. However, currently unclear is how the X-ray background of the ULX population, which represents the luminous end of the XRB luminosity function, evolves with redshift and how closely it follows that of typical XRBs.

To constrain this evolution, we first determine the complete overlapping area between the \CSCTwo~and the \sdss~footprints (Figure \ref{fig:SKY_MAP_CSC}). For each ULX candidate in our catalog, we determine the total solid angle of the \CSCTwo-\sdss~overlap with a \CSCTwo~0.5$-$7\,keV limiting sensitivity less than or equal to the ULX flux \citep[this corresponds to the total area over which the given ULX could have been detected; e.g.][]{Moretti:2003}. As a function of observed 0.5$-$7\,keV flux we compute the total number of sources (each weighted by the sky area over which it could have been detected). After correcting the numbers for spatial resolution limits (averaged over $L_{\mathrm{0.5-7,obs}}$\,$=$\,$10^{39}$\,$-$\,$2\times10^{41}$\,\uLum), contamination (Section \ref{sec:cont}), and galaxy inclination (Section \ref{sec:initmatch}), we then divide by the bin size to compute the differential distribution and parameterize it with a powerlaw function (Figure \ref{fig:dN_dS_OMEGA}). To obtain the total normalized flux, we integrate the quantity $SdN/dS/\Omega$ over the energy range for which we have measures of the differential distribution.

Figure \ref{fig:XRAY_Z_PLOT} shows the change in normalized flux per unit redshift of our sample (corrected for the mean fraction of hot ISM gas contribution for our sample in each redshift bin; see Section \ref{sec:confusion}). At low redshifts, our estimated value is consistent with that from the local sample of \citet{Swartz:2011}. We also show the estimated total point source emission from LMXBs and HMXBs based on X-ray detections and stacking (\citealp{Lehmer:2016}; assuming a Salpeter initial mass function and applying the stellar mass and \SFR~comoving density from \citealt{Madau:2014} and $k$-correcting to the observed 0.5$-$7\,keV energy range assuming an X-ray spectral index of $\Gamma=2.1$). Our results are consistent with the rising contribution from star forming galaxies as redshift increases and hence with ULXs being predominantly HMXBs. At \z\,$\sim$\,0.5, the ULXs are potentially consistent with accounting for all of the X-ray flux from HMXBs (i.e. due to star formation). We note, however, that when limiting our sample to \CSCTwo~sources with \texttt{likelihood\_class} values of TRUE and without any quality flags set (Section \ref{sec:final}), the contribution per unit redshift is lower by a factor of up to $\sim$\,2 (Figure \ref{fig:XRAY_Z_PLOT}).

The X-ray flux contribution per unit redshift increases out to \z\,$\sim$\,0.5 at a rate faster than predicted by the \SFR~comoving density. Population synthesis models predict that the evolution of XRB X-ray luminosities with \SFR~is driven by metallicity \citep[e.g.][]{Dray:2006,Linden:2010,Fragos:2013b,Madau:2017}, and recent observational studies have confirmed this \citep[e.g.][]{Douna:2015,Brorby:2016,Basu-Zych:2016,Ponnada:2020,Fornasini:2020}. Therefore, the stronger redshift evolution among our sample (relative to that of typical XRBs) may be caused by a stronger inverse dependence on metallicity, perhaps driven by larger stellar wind strength parameters \citep[e.g.][]{Belczynski:2010} associated with massive donor stars. Alternatively, an increasing association with more massive accretors at higher redshifts may also account for the additional flux. Regardless, at increasingly higher redshifts a larger fraction of normal galaxy point source X-ray emission is in the form of intrinsically luminous ULXs (or localized populations of ULXs) and reaches up to $\sim$\,40\%~at \z\,$\sim$\,0.5.

Figure \ref{fig:CXB_Z_PLOT} shows how the cumulative ULX contribution to the CXB evolves with redshift and how it compares to the XRB population of normal star-forming galaxies. When integrated out to \z\,$=$\,0.51, the total ULX contribution to the CXB is $\sim$\,\ULXCXBPercSumRnd\%~and consistent with the contribution from star forming galaxies. If the cumulative ULX contribution follows a similar redshift-dependent trajectory as that of HMXBs out to greater cosmological distances, then ULXs will account for at least $\sim$\,5\%~of the total CXB and hence contribute significantly to the overall ionizing flux from galaxies.

\section{Conclusions}
\label{sec:conc}

We present a catalog of \ULXSZ~ULX candidates covering the redshift range \z\,$=$\,\ZMin\,$-$\,\ZMax~that is constructed by matching \sdss~galaxies with the \ch~Source Catalog Version 2. After computing estimates of the relative astrometric accuracy between the \sdss~and \ch~images, off-nuclear X-ray sources are identified as those that are spatially offset from the host galaxy nucleus by $>$3 times the relative uncertainty in the X-ray and galaxy centroid positions. We further require that the unabsorbed, rest-frame 0.5$-$7\,keV~luminosities are above the commonly used lower threshold for ULXs ($10^{39}$\,\uLum) and below the upper threshold set by the most luminous known accreting compact stellar mass objects ($2\times10^{41}$\,\uLum). This catalog includes the largest sample of intermediate -redshift ULX candidates and extends out to higher redshifts than previous samples. We use the large redshift range of the catalog to constrain the physical nature of ULXs by analyzing the redshift evolution of their physical properties, occupation fractions, and contribution to the CXB. Our key conclusions are as follows:

\begin{itemize}

\item The overall contamination fraction is \ContFracwErrsb\%~and shows no evidence for a trend with redshift. Our estimates are consistent with those from previous nearby and intermediate-redshift samples when accounting for the uncertainties.

\item When comparing the ULX candidates to the parent sample, evidence for systematically enhanced \sSFRs~(at a given stellar mass) is observed. The median value of this offset is $\Delta \sSFR$\,$=$\,\MedDeltasSFR\,dex) and consistent with observations of local ULXs preferentially residing in or near star forming environments.

\item The \sSFRs~are statistically consistent with contributions from both LMXBs and HMXBs, though with a systematic bias toward a higher HMXB contribution. The ratios of X-ray luminosity to host galaxy stellar mass and \SFR~are consistent with XRB populations at low redshifts (agreement within \alphaSigOff$\sigma$ and \betaSigOff$\sigma$, respectively), but are significantly elevated toward higher redshifts (at the \gammaSigOff$\sigma$ and \deltaSigOff$\sigma$ levels, respectively). This deviation likely reflects an observational bias toward luminous sources that represent the extreme end of the XRB population and dominate their host galaxy X-ray emission.

\item The ULX occupation fraction, when corrected for source confusion, is positively correlated with redshift (per luminosity bin), as expected if ULX populations are preferentially found in galaxies with high \SFRs~and low metallicities. Our estimated occupation fractions are in agreement with previous results and show for the first time a systematic increase from the nearby Universe out to \z\,$\sim$\,0.5. 

\item The integrated contribution to the CXB from ULXs out to \z\,$=$\,0.51 is $\sim$\,\ULXCXBPercSumRnd\%. At these redshifts, the ULX X-ray background flux is consistent with that expected from HMXBs and accounts for up to $\sim$\,40\%~of the total normal galaxy X-ray point source flux. ULXs are therefore likely to contribute significantly to the overall ionizing radiation from galaxies.

\end{itemize}

\acknowledgments
{We thank an anonymous reviewer for the detailed and insightful comments that have greatly improved the manuscript quality. Support for this work was provided by NASA through \ch~Cycle 20 Proposal Number 20620227 issued by the \ch~X-ray Observatory Center, which is operated by the Smithsonian Astrophysical Observatory for and on behalf of NASA under contract NAS8-03060. The work of DS was carried out at the Jet Propulsion Laboratory, California Institute of Technology, under a contract with NASA. MH is supported by an ESO fellowship. This research has made use of data obtained from the \ch~Source Catalog, provided by the \ch~X-ray Center (CXC) as part of the \ch~Data Archive.}

\facilities{CXO, GALEX, PS1, Sloan, CTIO:2MASS, FLWO:2MASS, WISE}

\software{\astropy (\citealp{astropy:2013, astropy:2018}; \href{\astropylink}{\astropylink}).}

\end{document}